\documentclass[aps,pra]{revtex4-2}
\usepackage{siunitx}
\usepackage{booktabs}
\usepackage{amsmath}
\usepackage{graphicx} 
\usepackage{gensymb} 
\begin{document}
\title{Deriving the size and shape of the ALBA electron beam with optical synchrotron radiation interferometry using aperture masks: technical choices} \date{June 2024}

\author{Christopher L. Carilli}
\email[E-mail: ]{ccarilli@nrao.edu}
\affiliation{National Radio Astronomy Observatory, P. O. Box 0, Socorro, NM 87801, US}

\author{Laura Torino}
\email[E-mail: ]{ltorino@cells.es}
\affiliation{ALBA - CELLS Synchrotron Radiation Facility\\Carrer de la Llum 2-26, 08290 Cerdanyola del Vallès (Barcelona), Spain}

\author{Bojan Nikolic}
\email[E-mail: ]{bn204@cam.ac.uk}
\affiliation{Cavendish Laboratory, University of Cambridge, Cambridge CB3 0HE, UK}

\author{Ubaldo Iriso}
\email[E-mail: ]{uiriso@cells.es}
\affiliation{ALBA - CELLS Synchrotron Radiation Facility\\Carrer de la Llum 2-26, 08290 Cerdanyola del Vallès (Barcelona), Spain}

\author{Nithyanandan Thyagarajan}
\affiliation{Commonwealth Scientific and Industrial Research Organisation (CSIRO), Space \& Astronomy, P. O. Box 1130, Bentley, WA 6102, Australia}

\begin{abstract}
We explore non-redundant aperture masking to derive the size and shape of the ALBA synchrotron light source at optical wavelengths using synchrotron radiation interferometry. We show that non-redundant masks are required due to phase fluctuations arising within the experimental set-up. We also show, using closure phase, that the phase fluctuations are factorizable into element-based errors. We employ multiple masks, including 2, 3, 5, and 6 hole configurations. We develop a process for self-calibration of the element-based amplitudes (square root of flux through the aperture), which corrects for non-uniform illumination over the mask, in order to derive visibility coherences and phases, from which the source size and shape can be derived. We explore the optimal procedures to obtain the most reliable results with the 5-hole mask, based on the temporal scatter in measured coherences and closure phases. We find that the closure phases are very stable, and close to zero (within $2^o$). Through uv-modeling, we consider the noise properties of the experiment and conclude that our visibility measurements per frame are likely accurate to an rms scatter of $\sim 1\%$. 
\end{abstract}

\maketitle

\setlength{\parskip}{5pt}

\section{Introduction}

We consider the measurement of the ALBA synchrotron electron beam size and shape using optical interferometry with aperture masks. Monitoring the emittance of the electron beam is important for optimal operation of the synchrotron light source, and potentially for future improved performance and real-time adjustments. 

There are a number of methods to monitor the size of the electron beam, including: (i) LOCO, which is a guiding magnetic lattice analysis incorporating the beam position monitors, (ii) X-ray pinholes (Elleaume et al 1995), and (iii) Synchrotron Radiation Interferometry (SRI). Herein, we consider optical SRI, which can be done in real time without affecting the main beam. Previous measurements using SRI at ALBA have involved a two hole Young's slit configuration, with rotation of the mask in subsequent measurements to determine the two dimensional size of the electron beam, assuming a Gaussian profile (Torino \& Iriso 2016; Torino \& Iriso 2015). Such a two hole experiment is standard in synchrotron light sources (Mitsuhasi 2012; Kube 2007), and has been implemented at large particle accelerators, including the LHC (Butti et al. 2022). Four hole square masks have been considered for instantaneous two dimensional size characaterization, but such a square mask has redundant spacings which can lead to decoherence, and require a correction for variation of illumination across the mask (Masaki \& Takano 2003; Novokshonov et al. 2017; see Section~\ref{sec:redundancy}). Non-redundant masks have been used in synchrotron X-ray interferometry, but only for linear (one dimensional grazing incidence) masks (Skopintsev et al. 2014). 

In this report, we explore various configurations of a multi-hole, two-dimensional mask, emphasizing non-redundant masks, for an instantaneous measurement of the 2D electron beam size. Non-redundant masks have been used extensively in optical astronomical interferometric imaging to determine eg. the size of stellar photospheres, and for exoplanet searches (see Monnier 2003, Labeyrie 1996, Haniff et al. 1989), including recent observations with the aperture mask on the JWST (Hinkley et al. 2022; Lau et al. 2023). 

The beam size measurements will be presented in a parallel paper (Nikolic et al. 2024). The purpose of this report is to review the experimental setup, and discuss the adopted standard processing of the data for this approach to synchrotron light source size measurements. We then present the details of why various processing decisions were made, based on the experimental data.

In general, the ALBA SRI facility is ideal to explore various aspects of interferometry, including the effects of redundancy, shape-orientation-size conservation for three apertures (Thyagarjan \& Carilli 2022), and image plane self-calibration (Carilli, Nikolic, Thyagarjan 2023; Carilli et al. 2024). 

\section{Basics of Interferometric Imaging and Nomenclature}

The spatial coherence (or \textit{visibility}), $V_{ab}(\nu)$,
corresponds to the cross correlation of two quasi-monochromatic
voltages of frequency $\nu$ of the same polarization measured by two
spatially distinct elements in the aperture plane of an
interferometer. The visibility relates to the intensity distribution
of an incoherent source, $I(\hat{\mathbf{s}},\nu)$, via a Fourier
transform (van Cittert 1934, Zernike 1938, Born \& Wolf 1999; Thomson,
Moran, Swenson 2023):

\begin{align}
    V_{ab}(\nu) &= \int_\textrm{source} A_{ab}(\hat{\mathbf{s}},\nu) I(\hat{\mathbf{s}},\nu) e^{-i2\pi \mathbf{u}_{ab}\cdot \hat{\mathbf{s}}} \mathrm{d}\Omega \, , \label{eqn:VCZ-theorem}
\end{align}

\noindent where, $a$ and $b$ denote a pair of array elements (eg. holes in a mask), $\hat{\boldsymbol{s}}$ denotes a unit vector in the direction of any location in the image, $A_{ab}(\hat{\mathbf{s}},\nu)$ is the spatial response (the `power pattern') of each element (in the case of circular holes in the mask, the power pattern is the Airy disk), $\mathbf{u}_{ab}=\mathbf{x}_{ab} (\nu/c)$ is referred to as the ``baseline'' vector which is the vector spacing ($\mathbf{x}_{ab}$) between the element pair expressed in units of wavelength, and $\mathrm{d}\Omega$ is the differential solid angle element on the image (focal) plane. 

In radio interferometry, the voltages at each element are measured by phase coherent receivers and amplifiers, and visibilities are generated through subsequent cross correlation of these voltages using digital multipliers (Thomson, Moran, Swenson 2023; Taylor, Carilli, Perley 1999). In the case of optical aperture masking, interferometry is performed by focusing the light that passes through the mask (= the aperture plane element array), using reimaging optics (effectively putting the mask in the far-field, or Fraunhofer diffraction), and generating an interferogram on a CCD detector at the focus. The visibilities can then be generated via a Fourier transform of these interferograms or by sinusoidal fitting in the image plane. 

However, the measurements can be corrupted by distortions introduced by the propagation medium, or the relative illumination of the holes, or other effects in the optics, that can be described, in many instances, as a multiplicative element-based complex voltage gain factor, $G_a(\nu)$. Thus, the corrupted measurements are given by:
\begin{align}
    V_{ab}^\prime(\nu) &= G_a(\nu) \, V_{ab}(\nu) \, G_b^\star(\nu) \, , \label{eqn:uncal-vis}
\end{align}
where, $\star$ denotes a complex conjugation. 

The process of calibration determines these complex voltage gain factors. In general, calibration of interferometers can be done with one or more bright sources (`calibrators'), whose visibilities are accurately known (Thomson, Moran, Swenson 2023). Equation~(\ref{eqn:uncal-vis}) is then inverted to derive the complex voltage gains, $G_a(\nu)$ (Schwab1980, Schwab1981, Readhead \& Wilkinson 1978; Cornwell \& Wilkinson 1981). If these gains are stable over the calibration cycle time, they can then be applied to the visibility measurements of the target source, to obtain the true source visibilities, and hence the source brightness distribution via a Fourier transform.  

In the case of SRI at ALBA, we have employed self-calibration assuming a Gaussian shape for the synchrotron source, the details of which are presented in the parallel paper (Nikolic et al. 2024). Our process has considered only the gain amplitudes, corresponding to the square root of the flux through an aperture (recall, power $\rm \propto voltage^2$), dictated by the illumination pattern across the mask. We do not consider the visibility phases. Future work will consider full phase and amplitude self-calibration to constrain more complex source geometries.  

Closure phase is a quantity defined early in the history of astronomical interferometry, as a measurement of the properties of the source brightness distribution that is robust to element-based phase corruptions (Jennison 1958). Closure phase is the sum of three visibility phases measured cyclically on three interferometer baseline vectors forming a closed triangle, i.e., closure phase is the argument of the bispectrum = product of three complex visbilities in a closed triad of elements:

\begin{align}
    \phi_{abc}(\nu) &= \phi_{ab}(\nu) + \phi_{bc}(\nu) + \phi_{ca}(\nu) \, . \label{eqn:CPhase}
\end{align}

\noindent In this summation, the element-based phase errors cancel, and the measured closure phase equals the true closure phase, independent of calibration. Closure phase is image shift invariant, and it relates to the symmetry properties of the source (Section~\ref{sec:closure}). Closure phase is conserved under element-based complex gain calibration. 

Thyagarajan \& Carilli (2022)  present a geometric understanding of how closure phase manifests itself in the image plane. In essence, the shape, orientation, and size (SOS) of the triangle enclosed by the fringes of a three element interference pattern, are invariant to element-based phase errors, the only degree of freedom being an unknown translation of the grid pattern of triangles due to the element-based phase errors. A straight-forward means of visualizing how SOS conservation works is given in Figure 4 in Thyagarajan \& Carilli (2022): for any three element interferometer, the only possible image corruption due to an element-based phase screen is a tilt of the aperture plane, leading to a shift in the image plane. No higher order decoherence or image blurring is possible, since three points always define a plane parallel to which the wavefronts are coherent. This is not true for an image made with four or more elements, since multiple phase-planes can occur for different triads, and higher order decoherence (ie. image blurring) occurs. SOS conservation then raises the possibility of an image-plane self-calibration process (IPSC; Carilli, Nikolic, Thyagarajan 2023). We consider IPSC in a separate report (Carilli et al. 2024). 

\section{Experimental Setup}
\label{sec:setup}

The Xanadu optical bench setup at the ALBA synchrotron light source was the same as that used in Torino \& Iriso (2016), including aperture mask location, reimaging optics to achieve far-field equivalence, narrow band filters centered at 538~nm with a bandwidth of 10~nm, and CCD camera imaging. The distance from the mask to the target source, which is used to relate angular size measurements to physical size of the electron beam, was 15.05~m. The  optical extraction mirror is located 7~mm above the radiation direction (orbital plane of the electrons), at a distance of 7~m from the electron beam, implying an off-axis angle of $0.057^o$

\begin{figure}[!htb]
\centering 
\centerline{\includegraphics[scale=0.3]{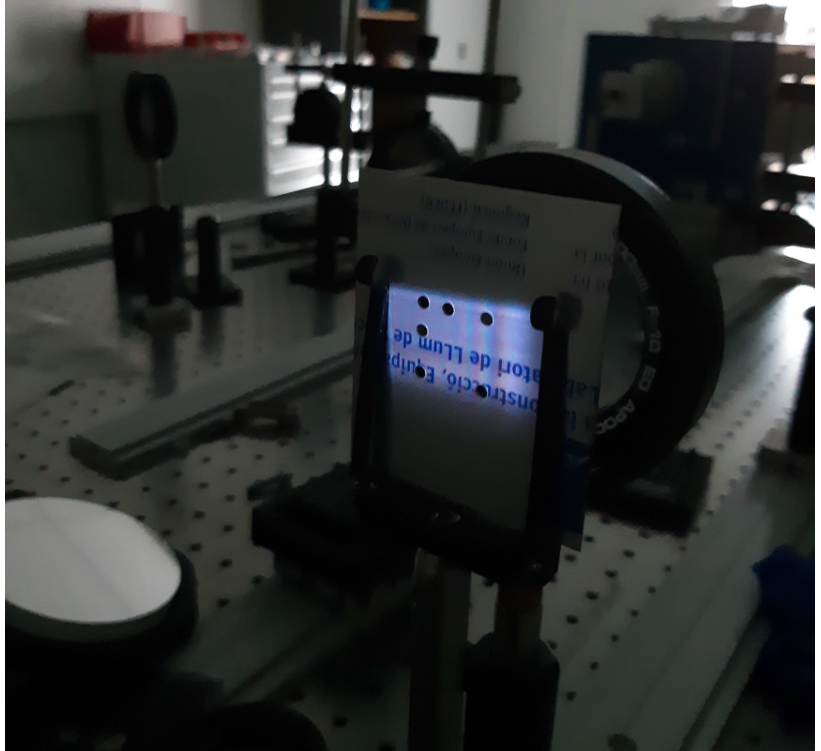}}
\caption{Full six hole mask on optical bench, with 3mm hole diameters. Combinations of holes in the top left corner were covered for the experiments. The hole numbering and geometry for the  mask is shown in Figure~\ref{fig:mask}.
}
\label{fig:mask}
\end{figure}

We employ multiple aperture masks. Figure~\ref{fig:mask} shows the full mask on the optical bench, with the illumination pattern from the synchrotron.  The full mask had 6 holes. Aperture masks of differing number of holes were generated by simply covering various holes for a given measurement. The geometry of the 6-hole mask is shown schematically in Figure~\ref{fig:mask}.  The mask was machined in the ALBA machine shop to a tolerance we estimate to be better than 0.1\,mm in hole position and size, based on measurements of the fringe spacings in the intensity images, and coordinates of the u,v points in the visibility plane. 

Note that the target source size is $\le 60\mu m$, which at a distance of 15.05\,m implies an angular size of $\le 0.84"$. For comparison, the angular interferometric fringe spacing of our longest baseline in the mask of 22.6\,mm at 540~nm wavelength is $5"$. This maximum baseline in the mask is dictated by the illumination pattern on the mask (Figure~\ref{fig:mask}). Hence, for all of our measurements, the source is only marginally resolved, even on the longer baselines. However, the signal to noise is extremely high, with millions of photons in each measurement, thereby allowing size measurements on partially resolving baselines. 

\begin{figure}[!htb]
\centering 
\centerline{\includegraphics[scale=0.25]{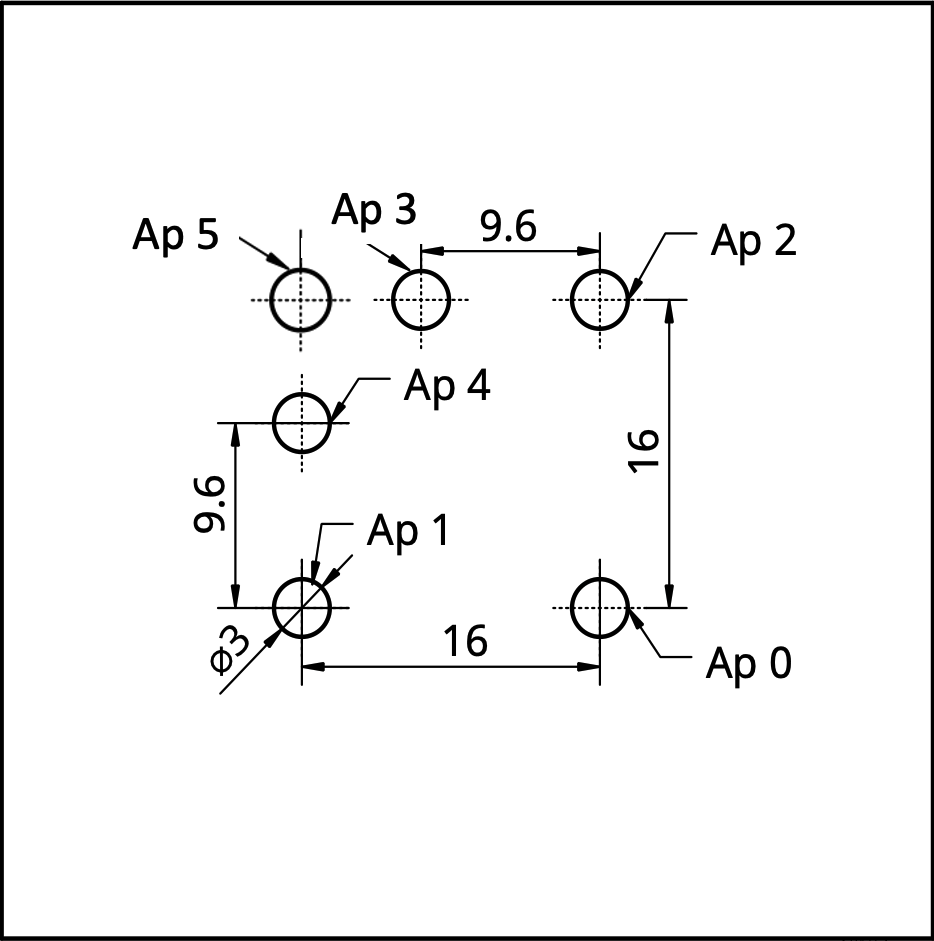}}
\caption{Scale drawing of the 6-hole mask with aperture labels as used for labelling baselines. Dimensions are in mm.}
\label{fig:mask6H}
\end{figure}

We consider masks with 2, 3, 5, and 6 holes. The 2-hole experiment employs a 16\,mm hole separation, and the mask is rotated by $45^o$ and $90^o$ sequentially to obtain two dimensional information, as per Torino \& Irison (2016). The 3-hole mask experiment employs apertures Ap0, Ap1, and Ap2. The 5-hole experiment used all of the apertures except Ap5 (see Figure~\ref{fig:mask6H}). 

In most cases, we employ non-redundant masks. A non-redundant aperture mask has a hole geometry such that each interferometric baseline, or separation between holes, is sampled uniquely in the Fourier domain (herein, called, the u,v plane), by a single pair of holes (Bucher \& Haniff 1993; Labeyrie 1996). Non-redundant masks are extensively used in astronomical interferometric imaging, in situations where the interferometric phases may be corrupted by atmospheric turbulence, or other phenomena that may be idiosyncratic to a given aperture (often referred to as 'element based phase errors').  In such cases, redundant sampling of a u,v point by multiple baselines with different phase errors would lead to decoherence of the summed fringes in the image (focal) plane. Similarly, a full aperture (ie. no mask), which has very many redundant baselines, will show blurring of the image due to this 'seeing' caused by phase structure across the aperture. Our 5-hole mask is an adaptation of Gonzales-Mejia (2011) non-redundant array, with the five holes selected to maximize the longer baselines, given the source is only marginally resolved.

The full 6-hole mask includes a square for the four corner holes, leading to two redundant baselines (horizontal 16~mm 2-5 and 0-1; vertical 16~mm 1-5 and 0-2). These will be used for testing of the effects of redundant sampling in Section~\ref{sec:redundancy}.

Masks were made with hole diameters of 3mm and 5mm, to investigate decoherence caused by possible phase fluctuations across a given hole. Observations were made with integration times (frame times) of 1\,ms and 3\,ms, to investigate decoherence by phase variations in time. Thirty frames are taken, each separated by 1 sec. 

We estimate the pixel size in the CCD referenced to the source plane of 0.138 arcsec/pixel, using the known hole separations (baselines), and the measured fringe spacings, either in the image itself, or in the Fourier transformed u,v distribution. 

\section{Standard Processing and Results}
\label{sec:processing}

\subsection{Images}

\begin{figure}[!htb]
\centering 
\centerline{\includegraphics[scale=0.25]{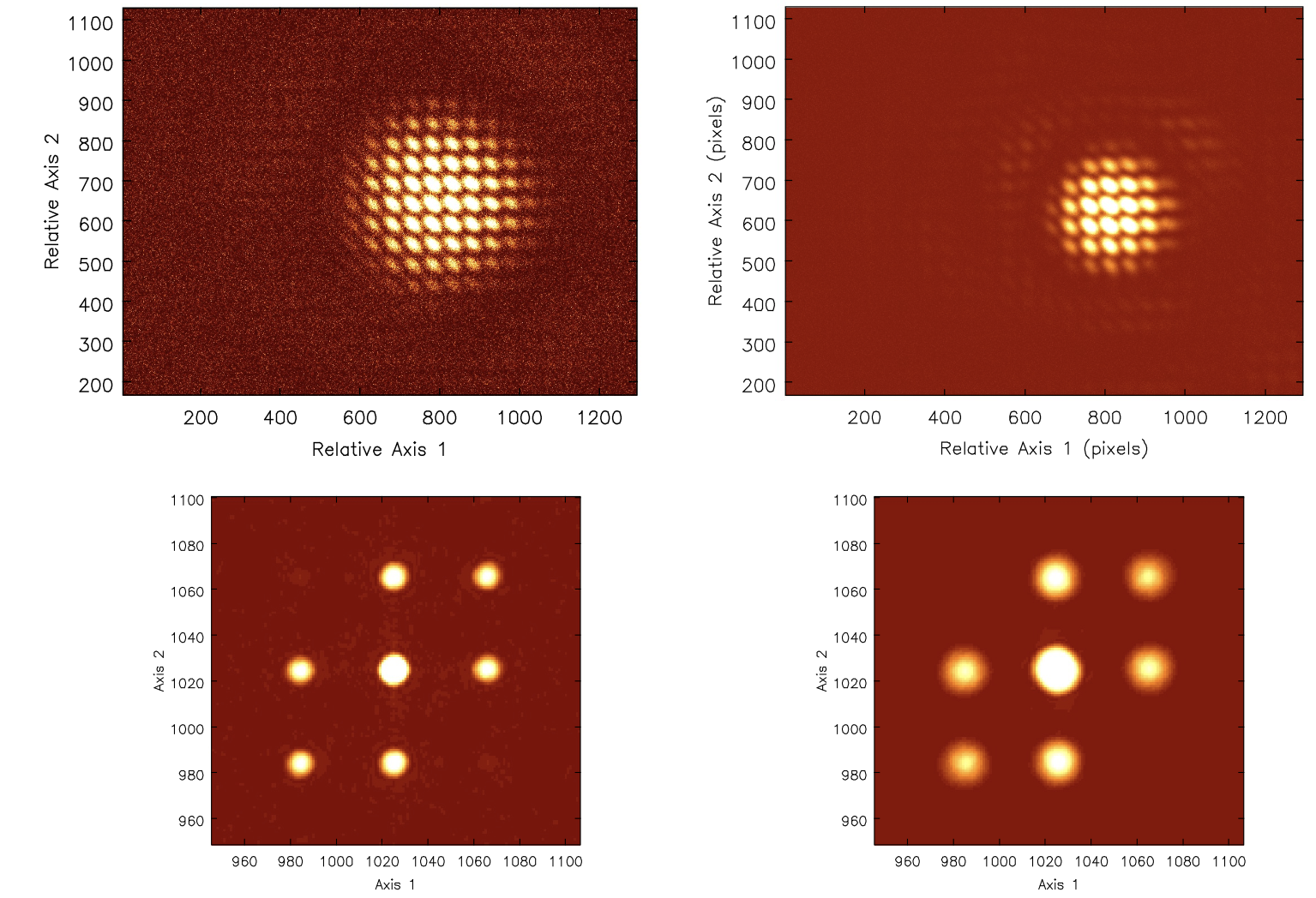}}
\caption{Upper frames: Example images for a three hole mask with 3mm (left) and 5mm (right) holes. Note how the size of the Airy disk (the overall intensity envelope), increases with decreasing hole size. Also seen are photons out beyond the first null in the Airy disk, in particular for the larger holes where there are consequently more photons. Lower frames: the Fourier transform of the two images (see Section~\ref{sec:FT}). Note how the convolution kernel in the Fourier domain decreases in size with increasing primary beam size, or decreasing aperture size.
}
\label{fig:PB}
\end{figure}

Figure~\ref{fig:PB} shows two images made with the 3-hole mask, one with 3\,mm holes and one with 5\,mm holes.  Any three hole image will show a characteristic regular grid diffraction pattern, modulated by the overall power pattern of the individual holes (Thyagarajan \& Carilli 2022). This power pattern envelope (the 'primary beam' for the array elements), is set by the hole size and shape, which, for circular holes with uniform illumination, appears as an Airy disk. The diameter of the Airy disk is $\propto \lambda/D$, where $\lambda$ is the wavelength and $D$ is the diameter. 

Also shown in Figure~\ref{fig:PB} are the Fourier transforms of the images (see Section~\ref{sec:FT}). The point here is that the size of the uv-samples decreases with increasing beam size = decreasing hole size. The primary beam power pattern (Airy disk) multiples the image-plane, which then corresponds to a convolution in the uv-plane. So a smaller hole has a larger primary beam and hence a smaller convolution kernel in the Fourier domain. 

These images show the expected behaviour, with the diffraction pattern covering more of the CCD for the 3\,mm hole image vs. the 5mm hole. Note that the total counts in the field is very large (millions of photons), and hence the Airy disk is visible beyond the first null, right to the edge of the field. This extent may be relevant for the closure phase analysis below. 

\begin{figure}[!htb]
\centering 
\centerline{\includegraphics[scale=0.25]{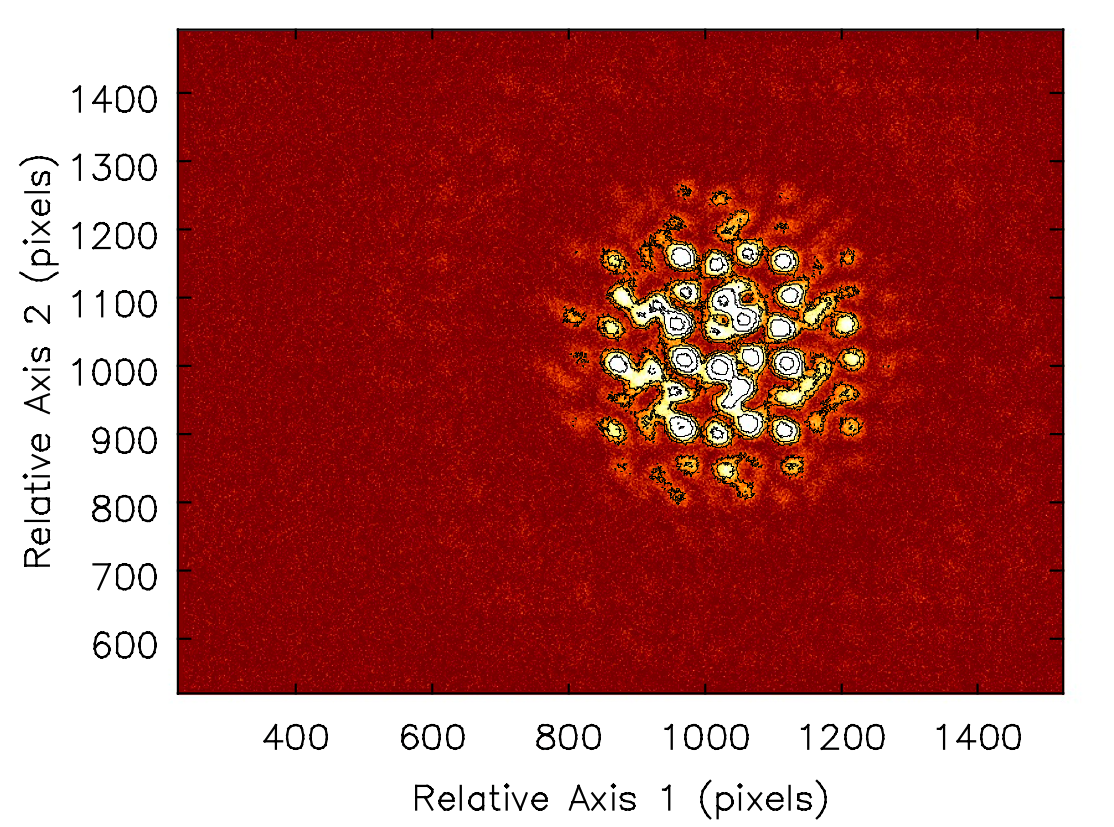}}
\caption{Typical interferogram for a 5-hole mask with 3mm-diameter holes and 1ms integration time. The CCD image is bias-subtracted. Contours are a geometric progression in factors of two, starting at 25 counts per pixel.}
\label{fig:3mminterf}
\end{figure}

Figure~\ref{fig:3mminterf} shows the corresponding image for a 5-hole mask with 3\,mm holes. The interference pattern is clearly more complex given the larger number of non-redundant baselines sampled ($\rm N_{baselines} = (N_{holes}*(N_{holes} - 1))/2$ = 10 for $\rm N_{holes} = 5$). 

\subsection{Fourier Domain}
\label{sec:FT}

Data are acquired as CCD two-dimensional arrays of size $1296\times 966$. We first remove the constant offset which is due to a combination of the bias and the dark current. We use a fixed estimate of this offset obtained by examination of the darkest areas of the CCD and the FFT of the image. We find a bias of 3.7 counts per pixel. Errors in this procedure accumulate in the central Fourier component, corresponding to the zero spacing, or total flux (u,v = 0,0), and contribute to the overall uncertainty of the beam reconstruction. 

Next we pad and center the data so that the centre of the Airy disk-like envelope of the fringes is in the centre of a larger two-dimensional array of size $2048 \times 2048$. To find the correct pixel to center to we first smooth the image with a wide (50 pixel) Gaussian kernel, then select the pixel with highest signal value. The Gaussian filtering smooths the fringes creating an image corresponding approximately to the Airy disk. Without the filtering the peak pixel selected would be affected by the fringe position and the photon noise, rather than the envelope. Off-sets of the Airy disk from the image center lead to phase slopes across the u,v apertures. 

\begin{figure}[!htb]
\centering 
\centerline{\includegraphics[scale=0.3]{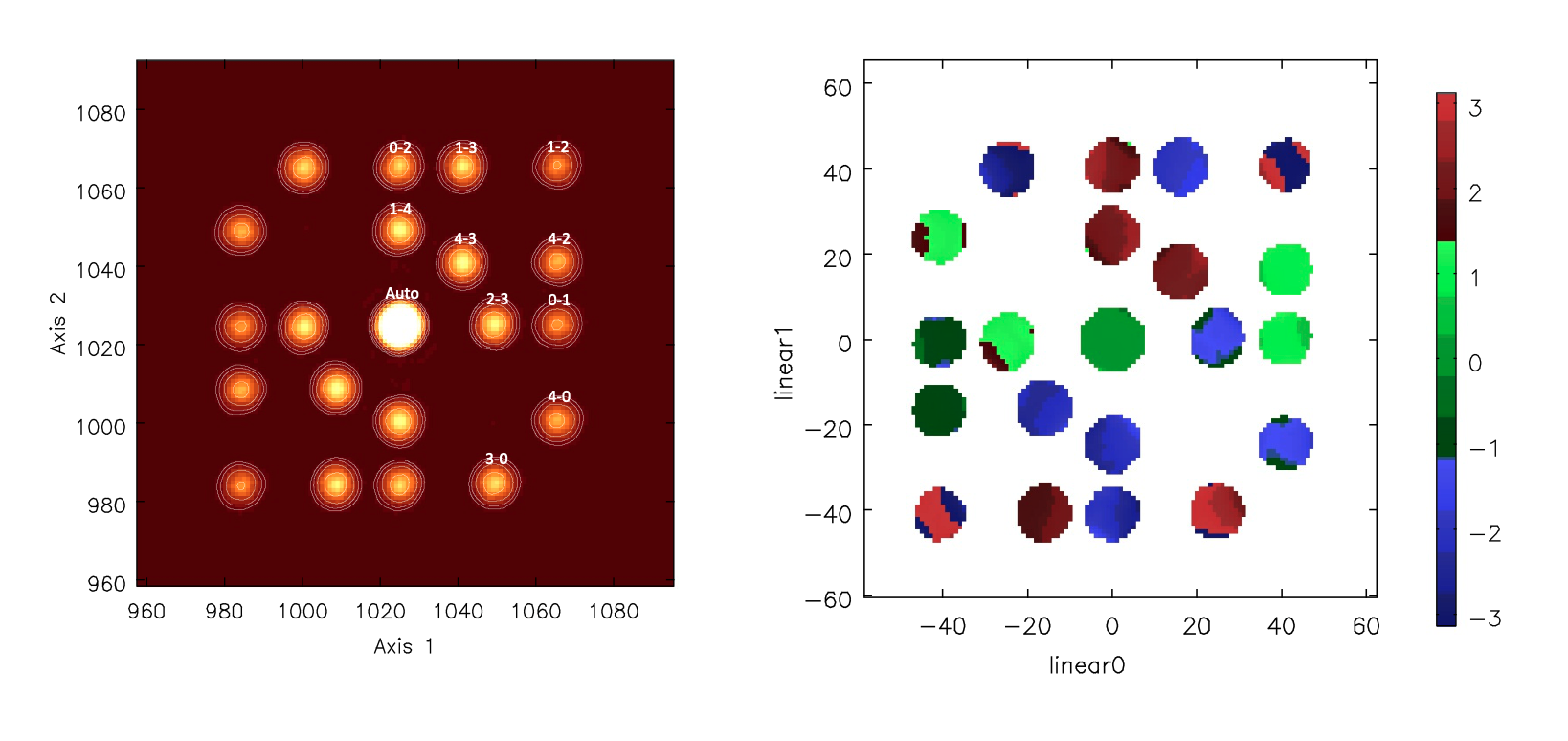}}
\caption{Amplitude and phase of the FFT of the example 5-hole interferogram in  Figure~\ref{fig:3mminterf}.  Contours are a geometric progression in factor two. Labels identify peaks of amplitude with the pair of apertures forming the baseline of the visibility that the peak represents (Figure~\ref{fig:mask6H}). Phase units are radians.}
  \label{fig:3mmfft}
\end{figure}

To calculate the coherent power between the apertures, we make use of the van Cittert–Zernike theorem that the coherence and the image intensity are related by a Fourier transform.  We therefore compute the two-dimensional Fourier transform of the padded CCD frame using the FFT algorithm. Amplitude and phase images of an example Fourier transform are shown in Figure~\ref{fig:3mmfft}. Distinct peaks can be seen in the FFT corresponding to each vector baseline defined by the aperture separations in the mask.

We extract the correlated power on each of the baselines by calculating the complex sum of pixels within a circular aperture of 7 pixels, centered at the calculated position of the baseline. With the padding used here 1\,mm on the mask corresponds to 2.54 pixels in the Fourier transformed interferogram. An illustration of this procedure on the example frame is shown in Figure \ref{fig:3mmsection}. We experimented with different u,v apertures (3,5,7,9 pixels), and found that 7 pixels provided the highest S/N while avoiding overlap with the neighboring u,v sample (Section~\ref{sec:Npix}). 

\begin{figure}[!htb]
\centering 
\centerline{\includegraphics[scale=0.25]{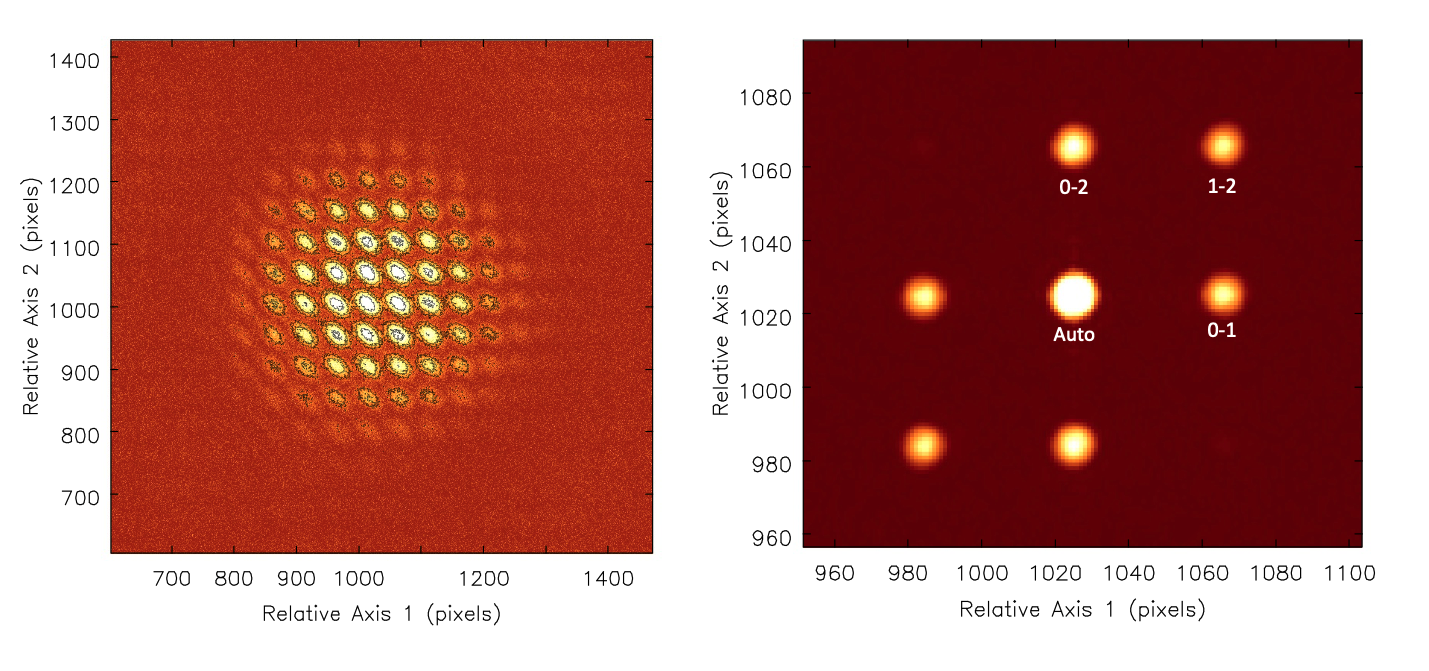} }
\caption{Intensity image and amplitudes of the visibilities for the three hole mask. Contours are a geometric progression in factors of two starting at 20 counts per pixel.}
  \label{fig:3H}
\end{figure}  

\begin{figure}[!htb]
\centering 
\centerline{\includegraphics[scale=0.25]{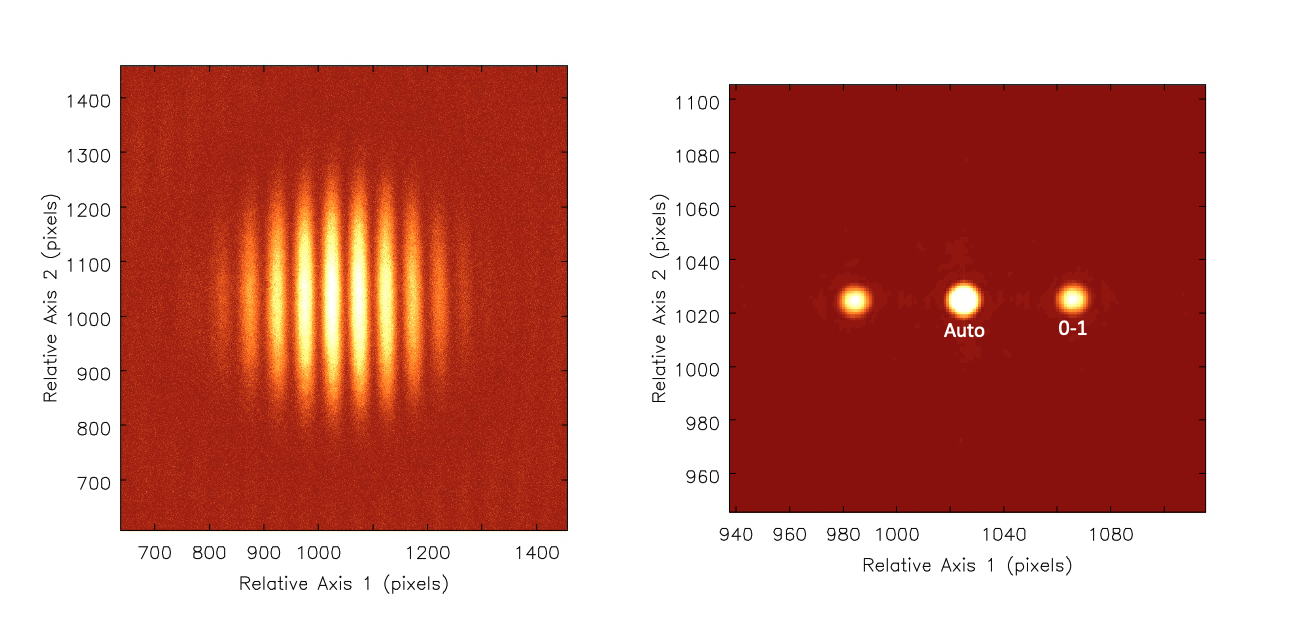}}
\caption{Intensity image and amplitudes of the visibilities for the two hole mask. }
  \label{fig:2H}
\end{figure}  

\begin{figure}[!htb]
\centering 
\centerline{\includegraphics[scale=0.25]{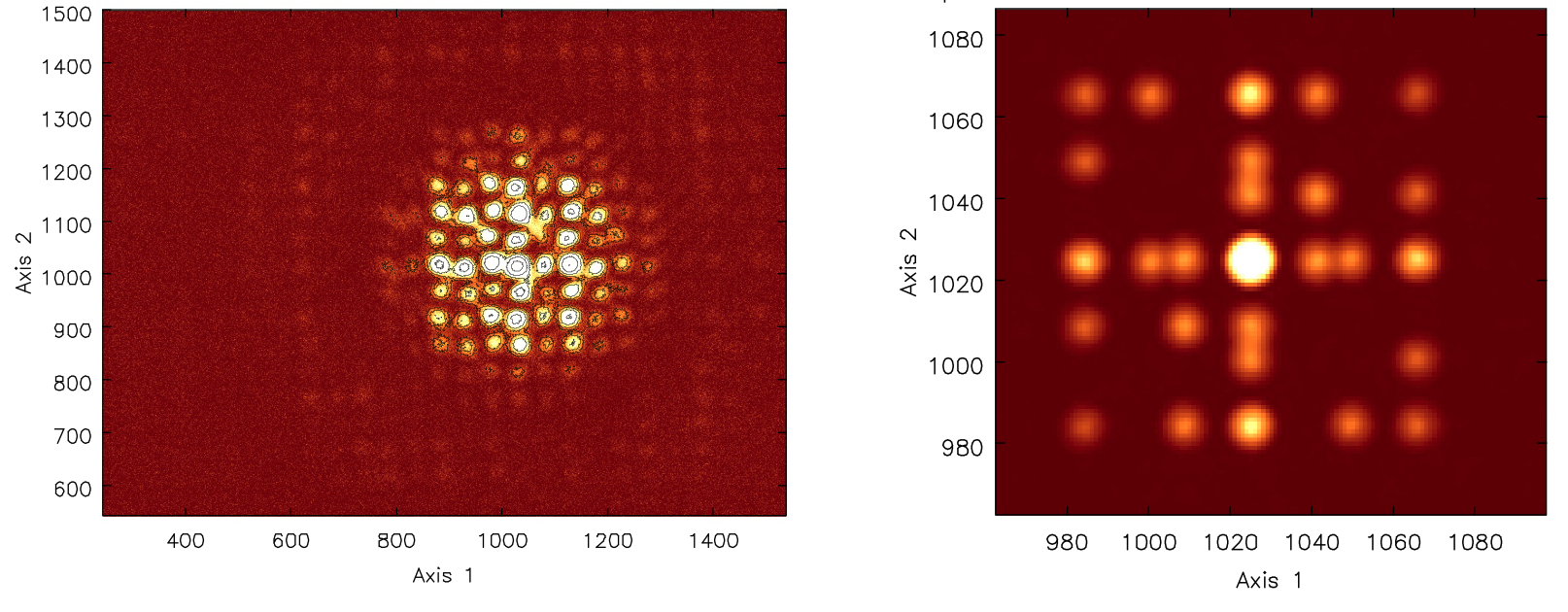}}
\caption{Intensity image and amplitudes of the visibilities for the six hole mask. Contour levels are a geometric progression in factors of two, starting at 25 counts per pixel.}
  \label{fig:6H}
\end{figure}  

The interferometric phases of the visibilities are derived by a vector average over the selected apertures in the uv-plane of the images of the Real and Imaginary part of the Fourier transform, using the standard relation: phase = arctan(Im/Re).

For reference, Figure~\ref{fig:3H} shows the intensity image and visibility amplitudes for a three hole mask with 3\,mm holes and 1\,ms integrations,  Figure~\ref{fig:2H} shows the same for one of the 2-hole mask with 3\,ms integrations, and Figure~\ref{fig:6H} shows the same for the 6-hole mask and 1\,ms integrations. The u,v pixel locations of the Fourier components are dictated by the mask geometry (ie. the Fourier conjugate of the hole separations or 'baselines`), and determined by the relative positions of the peaks of the sampled u,v points to the autocorrelation. These are set by the sampled baselines in the mask, the Fourier conjugate of which are the spatial frequencies. We find that the measured u,v data points are consistent with the mask machining to within 0.1\,mm, and that the u,v pixel locations for the common u,v sampled points between the 2-hole, 3-hole, and 5-hole mask agree to within 0.1 pixel. 

Notice that, for the 6-hole mask Figure~\ref{fig:6H}, the u,v data points corresponding to the vertical and horizontal 16mm baseline have roughly twice the visibility amplitude as neighboring points (and relative to the 5-hole mask). This is because these are now redundantly sampled, meaning the 16mm horizontal baseline now includes photons from 0-1 and 2-5, and 16mm vertical baseline includes 0-2 and 1-5. 

\subsection{Self-calibration}
\label{sec:selfcal}


The self-calibration and source size fitting is described in more detail in Nikolic et al. (2024). For completeness, we summarize the gain fitting procedure and equations herein, since it is relevant to the results presented below. 

For computational and mathematical convenience (see Nikolic et al. 2024), the coherence is modelled as a two-dimensional Gaussian function parametrised in terms of the overall width ($\sigma$) and the distortion in the `+` ($\eta$) and `X' ($\rho$) directions. Dispersion in e.g., the $u$ direction is $\sigma/\sqrt{1+\eta}$ while in the $v$ direction it is $\sigma/\sqrt{1-\eta}$, which shows that values $\eta$ or $\rho$ close to 1 indicate that one of the directions is poorly constrained.

\begin{align}
  \gamma(u,v) &=&\exp[- \frac{(u^2 + v^2) + 2 \rho (u v) + \eta(u^2 - v^2)}{2 \sigma^2} ]\\
  \|V_{ij}\| & = & \gamma(u_{ij}, v_{ij}) \|G_i\| \|G_j\|\\
  \|V_{\rm auto}\| & = & \sum_i  \|G_i\|^2
\end{align}

\begin{figure}[!htb]
\centering 
\centerline{\includegraphics[scale=0.5]{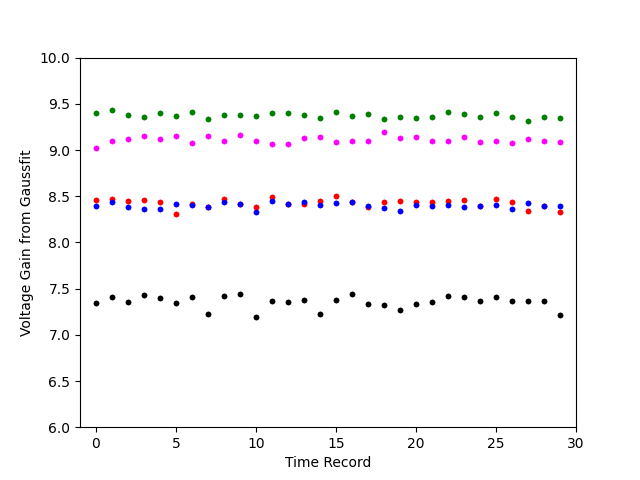}}
\caption{Derived voltage gains for the 5-hole mask.  The typical RMS deviation over time is $\le$ 1\%. 
}
\label{fig:gains5H}
\end{figure}

The fitting to the data is done using the Levenberg-Marquardt algorithm. The derived gains from the self-calibration process are shown in Figure~\ref{fig:gains5H}, and listed in Table~\ref{fig:gains}.

\begin{figure}[!htb]
\centering 
\centerline{\includegraphics[scale=0.3]{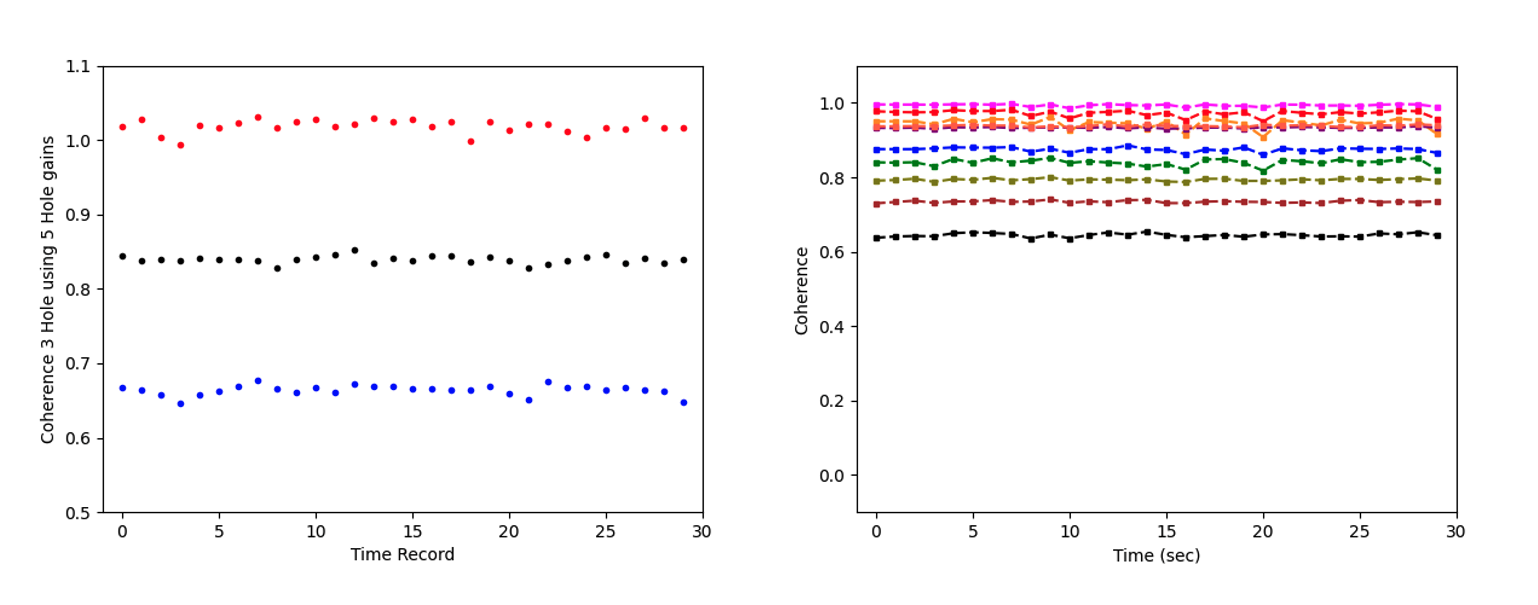}}
\caption{Plots of the coherence derived for the 3-hole and 5-hole data vs. time. The 3-hole coherences were derived using the 5-hole gains to correct for illumination. Values are listed in Table~\ref{fig:cohtab}.
}
\label{fig:3H5Hcoh}
\end{figure}

The derived coherences, ie. visibility fractional amplitudes (relative to the zero spacing) corrected for the voltage gains, are show in Figure~\ref{fig:3H5Hcoh}, for both 3-hole and 5-hole data, and listed in Table~\ref{fig:cohtab}. Gain fitting is not possible with 3-hole data since the problem becomes under-constrained, so the three hole coherences were derived assuming the 5-hole gains for illumination correction. The coherences are remarkably stable, to less than a percent. The 3-hole coherences are a few percent higher than the 5-hole data for matching baselines. This difference may relate to the assumed gains and possible different mask illumination for that particular experiment. Indeed, the fact that for one baseline on the 3-hole data some of the coherences are slightly larger than unity (which is unphysical), suggests the assumed gains may not be quite correct. A general point is that all the coherences are high, $\ge 65\%$, indicating that the source is only margninally resolved spatially. 

\begin{figure}[!htb]
\centering 
\centerline{\includegraphics[scale=0.6]{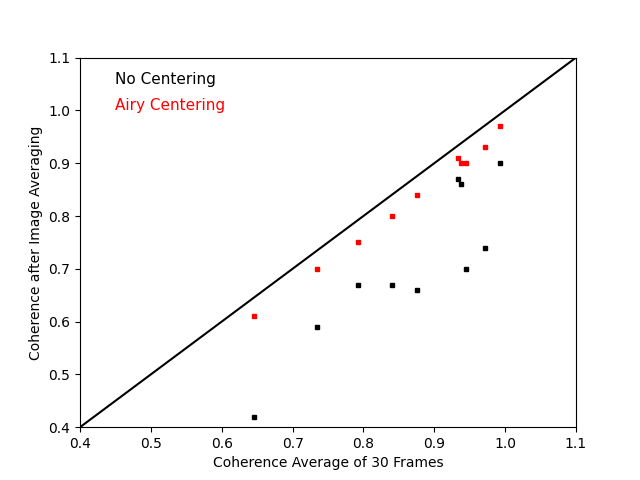}}
\caption{The horizontal axis shows the coherences for the 10
  visibilities for data from a 5-hole mask with 3\,mm holes, 1\,ms
  frame time, equal to the average value of the individual measurements for each of the 30 frames (see Table~\ref{fig:cohtab}). The vertical  axis shows the coherences derived after averaging the 30 images and  then determining the coherences. The black points are for images  averaged without any centering, and the red points including Airy
  disk centering. The solid line is equal
  coherence. All the points fall below this line, indicating that
  averaging the images before transforming leads to decoherence, with
  a larger affect involving no centering prior to image averaging.}
\label{fig:Coh5Have}
\end{figure}

Columns 6 and 7 in Table~\ref{fig:cohtab} list the coherences derived after summing the 30 images taken over the 30 seconds of the time series before deriving the coherences. These data are plotted in Figure~\ref{fig:Coh5Have}. If no centering at all is performed (column 6), the image wander and structural changes due to phase fluctuations across the field leads to decoherence up to 35\%. If image centering is performed based on the centering of the Airy disk before summing the images, the coherences increase, but remain lower than the average from each individual frame by 5\% or so. Centering on the Airy disk, corresponds effectively to a 'tip-tilt' correction, meaning correcting for a uniform phase gradient across the mask (the lowest order term in the phase screen). Decoherence is even seen when comparing 1\,ms to 3\,ms averaging (see Section~\ref{sec:tave}). 

Table~\ref{fig:gains} also lists the gains derived after image averaging, with and without Airy disk centering. In this case, the gains are essentially unchanged (within 1\%), relative to the mean from the time series (row 1). This similarity for gain results from data that clearly involved decoherence of the visibilities themselves lends confidence that the derived illumination correction (the 'gains'), are correct. 

\begin{table}
\begin{tabular}{rllllllll}
\toprule
 &  $G_0$ & $G_1$ & $G_2$ & $G_3$ & $G_4$ & $\sigma/ {\mathrm{cells}}$ & $\rho$ & $\eta$  \\
\midrule
  Mean of best-fits in time series & 7.35 & 8.43 & 8.40 &  9.37 &  9.11 &  74.66 &   0.9 &   0.66\\
  RMS in time series & 0.067 & 0.045 & 0.029 & 0.027 & 0.034 &  4.88 &  0.42& 0.13\\\midrule
  Sum of 30 frames with no centering &  7.45 & 8.50 & 8.47 & 9.35 & 9.15 & 49.1 & 0.22 & 0.15\\
  Sum of 30 frames with Airy centering & 7.41 & 8.45 & 8.50 & 9.37 & 9.13 & 68.8 & 0.37 & 0.58\\
\bottomrule
\end{tabular}
\caption{
Gains derived from the self-calibration process for a 5-hole mask.}
\label{fig:gains}
\end{table}

\begin{table}
  \centering 
\begin{tabular}{lcccccc}
\toprule
Baseline & ~~~5-hole Coherence & RMS & ~~~3-hole Coherence & RMS & No Center & Airy Center  \\
  \midrule
0-1 &  0.793 & 0.0030 & 0.816 & 0.0050 & 0.67 & 0.75\\
0-2 & 0.972 & 0.0079 & 0.989 & 0.0088  & 0.74 & 0.93 \\
0-3 & 0.945 & 0.0130 & ~ & ~ & 0.70  & 0.90 \\
0-4 & 0.840& 0.0089 & ~ & ~ & 0.67  & 0.80\\
1-2 & 0.645 & 0.0048 & 0.691 & 0.0073 & 0.42 & 0.61 \\
1-3 & 0.875& 0.0056 & ~ & ~  & 0.66 &  0.84\\
1-4 & 0.993 & 0.0030 & ~ & ~ & 0.90 &  0.97\\
2-3 & 0.933 & 0.0014 & ~ & ~ & 0.87 &  0.91\\
2-4 & 0.734 & 0.0028 & ~ & ~ & 0.59 &  0.70\\
3-4 & 0.938 & 0.0023 & ~ & ~ & 0.86 &  0.92\\
\bottomrule
\end{tabular}
\caption{Column 2 and 3: mean coherences and RMS scatter for the time series of measurements for the 10 baselines in the 5-hole mask, after Airy centering.  Column 4 and 5 lists the same for the 3-hole data. Column 6 lists the coherence derived by first summing all of the frames together, then doing the Fourier transform, without any image centering. Column 7 lists the same but after Airy disk centering. }
\label{fig:cohtab}
\end{table}

\subsection{Visibility and Closure Phases}
\label{sec:closure}

\begin{figure}[!htb]
\centering 
\centerline{\includegraphics[scale=0.25]{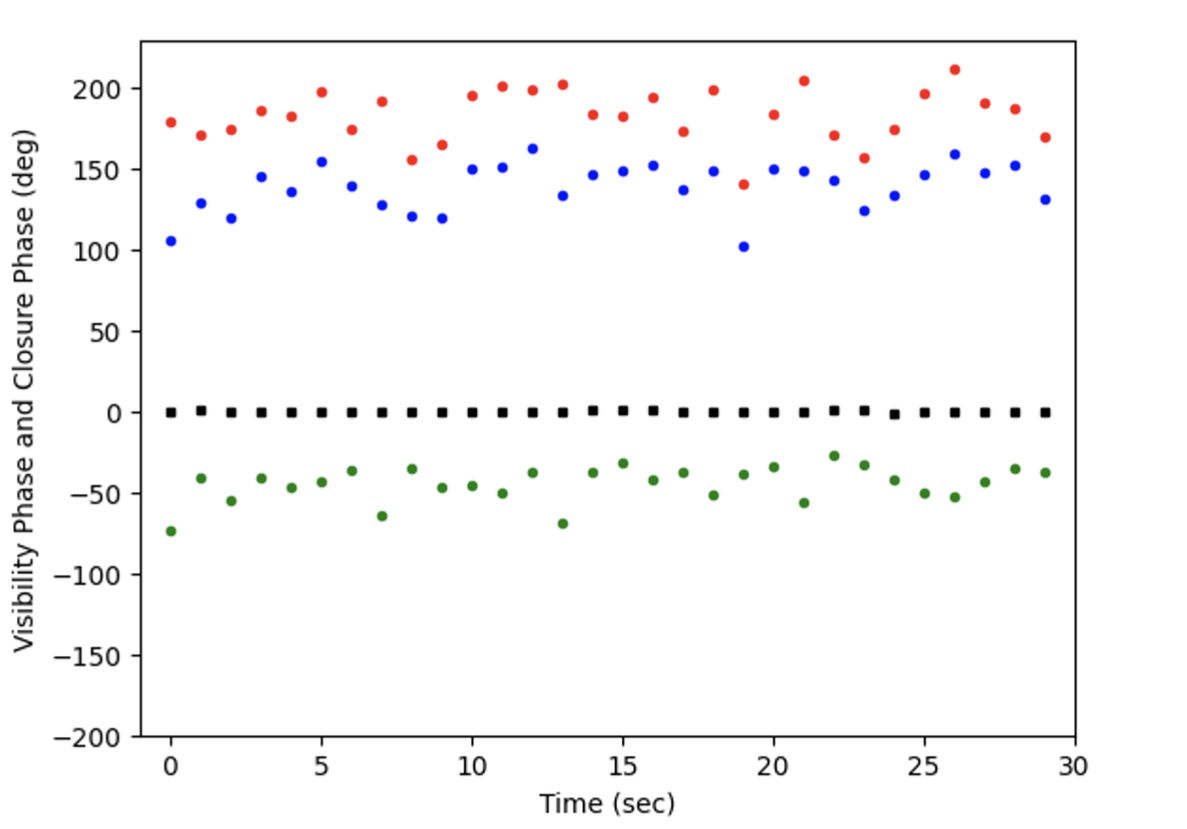}}
\caption{Visibility and closure phase on one closed triad of Fourier samples (holes 1,2,3) for the 5-hole mask, after Airy centering. Black: Closure phase for triad 1-2-3, with mean = $0.46^o$ and RMS over the time series of $0.49^o$. Blue: visibility phase for baseline 1-3 with mean = $139.4^o$ and RMS = $15^o$;  Green visibility phase for baseline 2-3 with mean = $-43.8^o$ and RMS = $11^o$;  Red: visibility phase for baseline 1-2 with mean = $183.7^o$ and RMS = $16^o$; }
\label{fig:visclph}
\end{figure}

Figure~\ref{fig:visclph} shows the measured visibility phases on three baselines (baseline 1-2, 2-3, 1-3; Figure~\ref{fig:mask}) as a function of time. There is substantial variation between the frames, with a peak to peak variation of about $50^o$. Such a variation would lead to decoherence (smearing) of the source if either (i) no aperture mask were present and a direct image were made, or (ii) a redundant mask was used, for which Fourier samples using redundant spacings would have lower apparent coherence due to independent phase fluctuations. 

Another interesting feature is that the larger phase fluctuations appear to correlate in time between baselines 1-2 and 1-3. These are the two longer baselines, with similar geometry (see Figure~\ref{fig:mask}). The correlation in time implies some spatial coherence of the phase screen across the full mask. In particular, a Kolmogorov turbulent power spectrum for phase fluctuations in the lab atmosphere (larger fluctuations on longer baselines), could lead to the measured correlation on the two long and similar baselines, although there may be other explanations, such as vibration of the optical elements (Torino \& Iriso 2015).

Figure~\ref{fig:visclph} also shows the `closure phase' for this triad of baselines (Equ.~\ref{eqn:CPhase}). The closure phase is remarkably constant, with an RMS deviation of only $0.49^o$, and a mean value very close to zero. The stability of the closure phase in time, compared to the unstable visibility phases, implies that the phase fluctuations can be factored into aperture-based phase perturbations of the electromagnetic fields.

\begin{figure}[!htb]
\centering 
\centerline{\includegraphics[scale=0.27]{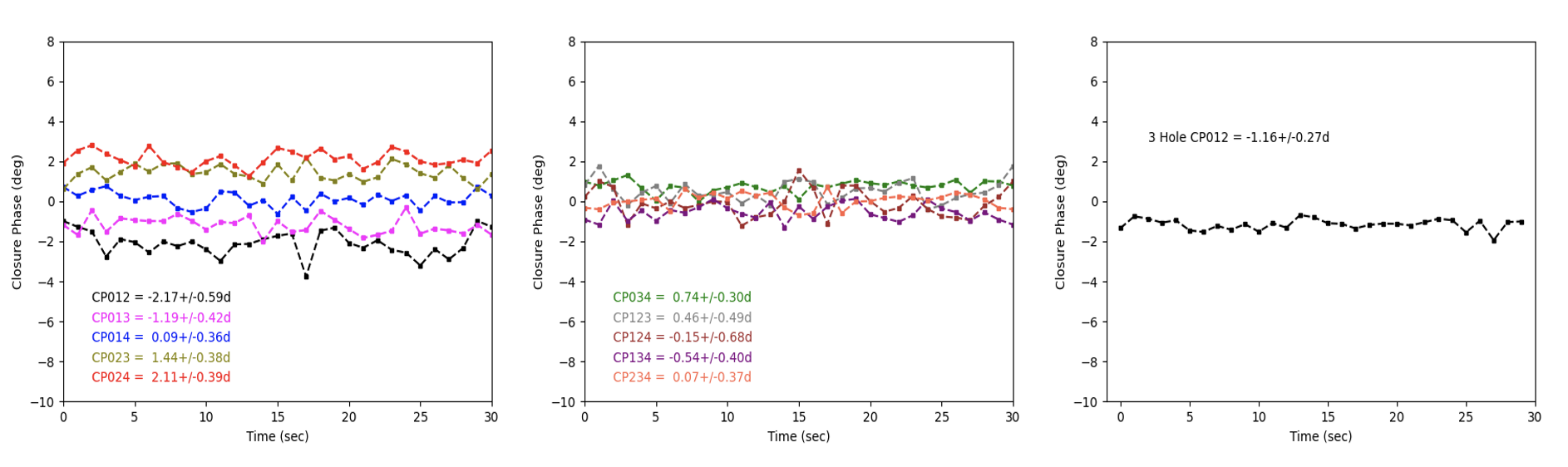}}
\caption{Left and Middle: Closure phases for the 10 triads in the 5-hole mask with 3\,mm holes, 1ms averaging. Values are listed in Table~\ref{fig:closureph}. Same, but now for 3-hole data with same processing. 
}
\label{fig:CLPH5H}
\end{figure}

Figure~\ref{fig:CLPH5H} shows the closure phases for all ten triads in the uv-sampling, and the values are listed in Table~\ref{fig:closureph}. All the closure phases are stable (RMS variations $\le 0.7^o$), and all the values are close to zero, typically $\le 1^o$. The only triads with closure phases of about $2^o$ involve the baseline 0-2. This is the vertical baseline of 16\,mm length, and hence has a fringe that projects (lengthwise) in the horizontal direction. The origin of closures phases that appear to be very small, but statistically different from zero, is under investigation. For the present, we conclude the closure phases are $< 2^o$.
 
Closure phase is a measure of source symmetry. X-ray pin-hole measurements imply that the beam is Gaussian in shape to high accuracy (Elleaume et al. 1995). A closure phase close to zero is typically assumed to imply a source that is point-symmetric in the image plane (a closure phase $\le 2^o$ implies  brightness asymmetries $\le 1\%$ of the total flux, for a well resolved source), as would be the case for an elliptical Gaussian. However, the fact that the source is only marginally resolved (Section~\ref{sec:setup}), can also lead to small closure phases, regardless of source structure on scales much smaller than the resolution. A simple test using uv-data for a very complex source that is only marginally resolved, shows that for closed triads composed of baselines with coherences $\ge 70\%$, the closure phase is $< 2^o$. In this case, even small, but statistically non-zero, closure phases provide information on source structure. 

\begin{table}
  \centering 
\begin{tabular}{lcc}
\toprule
Triad & ~~~Mean Closure Phase & RMS \\
~ & degrees & degrees \\
  \midrule
0-1-2 & -2.17 & 0.59 \\
0-1-3 & -1.19 & 0.42 \\
0-1-4 & 0.09 & 0.36 \\
0-2-3 & 1.44 & 0.38 \\
0-2-4 & 2.11 & 0.39 \\
0-3-4 & 0.74 & 0.30 \\
1-2-3 & 0.46 & 0.49 \\
1-2-4 & -0.15 & 0.68 \\
1-3-4 & -0.54 & 0.40  \\
2-3-4 & 0.07 & 0.37 \\
\bottomrule
\end{tabular}
\caption{Closure phases for the 10 triads in the 5-hole mask.}
\label{fig:closureph}
\end{table}

Figure~\ref{fig:CLPH5H} also shows the closure phase for the 3-hole data, which has only one triad (holes 0-1-2).  The 3-hole closure phase for triad 0-1-2 has a mean of $-1.16^o$ with an RMS of the time series of 0.27$^o$. For comparison, the values for the 5-hole data for this triad were $-2.17^o$ and $0.59^o$. The values ought to be the same, all else being equal.  The difference could arise from: 
(i) the source changed (unlikely),
(ii) the geometry of the mask changed (could only be a rotation): but the uv-sampling points found by photom are within 0.1 pixels,
(iii) the centering of the diffraction pattern on the CCD is different, which leads to a different sampling of the outer Airy disk. We are investigating these phenomena. For now, we can conclude is that $\sim 2^o$ is the limit to a reliable closure phase measurement from experiment to experiment, for the current data.

\section{Processing choices}

The analysis presented herein is meant as supporting material for other papers that present the science results. Our main focus is to justify the choices made in this new type of analysis of laboratory optical interferometric data. 

\subsection{Centering: phase slopes}

\begin{figure}[!htb]
\centering 
\centerline{\includegraphics[scale=0.2]{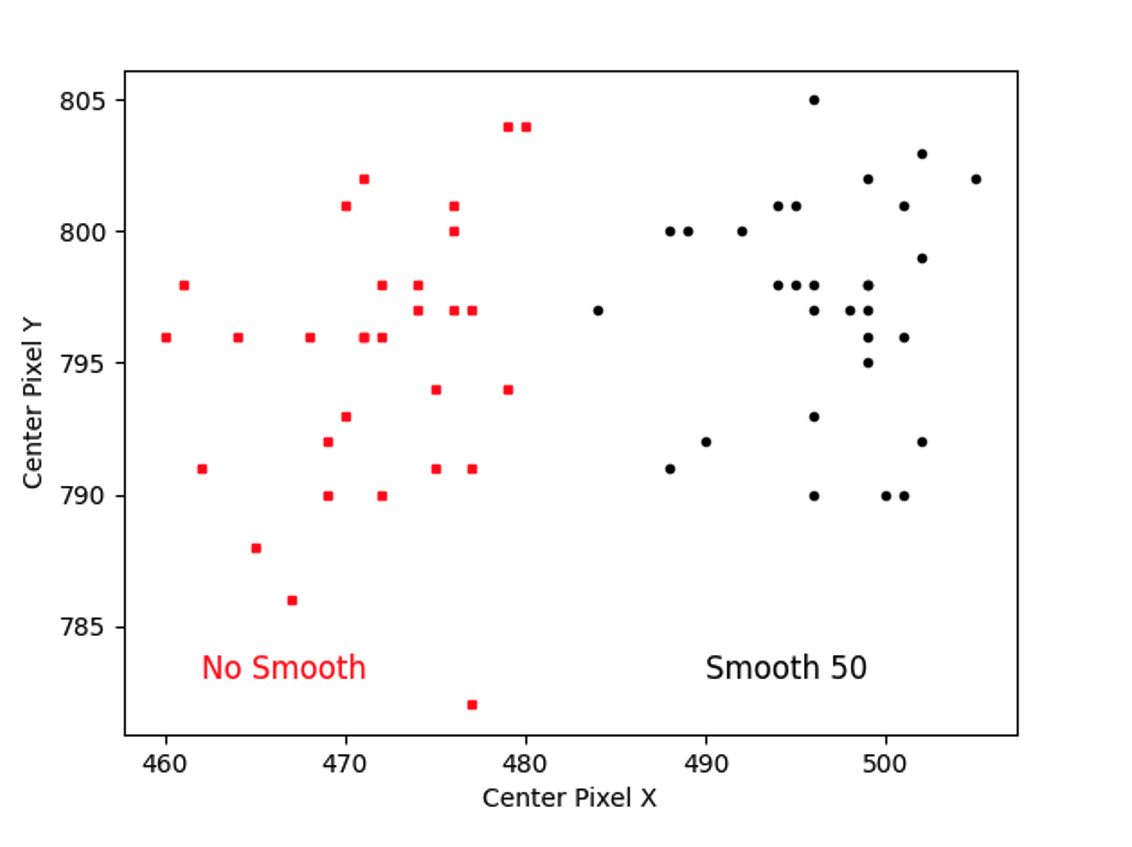}}
\caption{Scatter plot of the centering pixel for the 30 time records as reported by edfconvert with and without Airy disk centering. 
}
\label{fig:scatcent}
\end{figure}

For reference, Figure~\ref{fig:scatcent} shows the centers found with and without smoothing of the input image. Centering will affect mean phases and phase slopes across apertures. We have found that smoothing before centering, ie. centering on the Airy disk not the peak pixel, leads to the minimum phase slopes across the u,v sampled points, as seen in Figure~\ref{fig:3mmfft}. The scatter plot shows similar overall scatter with and without Airy disk centering, but there is a systematic shift, which leads to phase slopes across apertures.

\begin{figure}[!htb]
\centering 
\centerline{\includegraphics[scale=0.5]{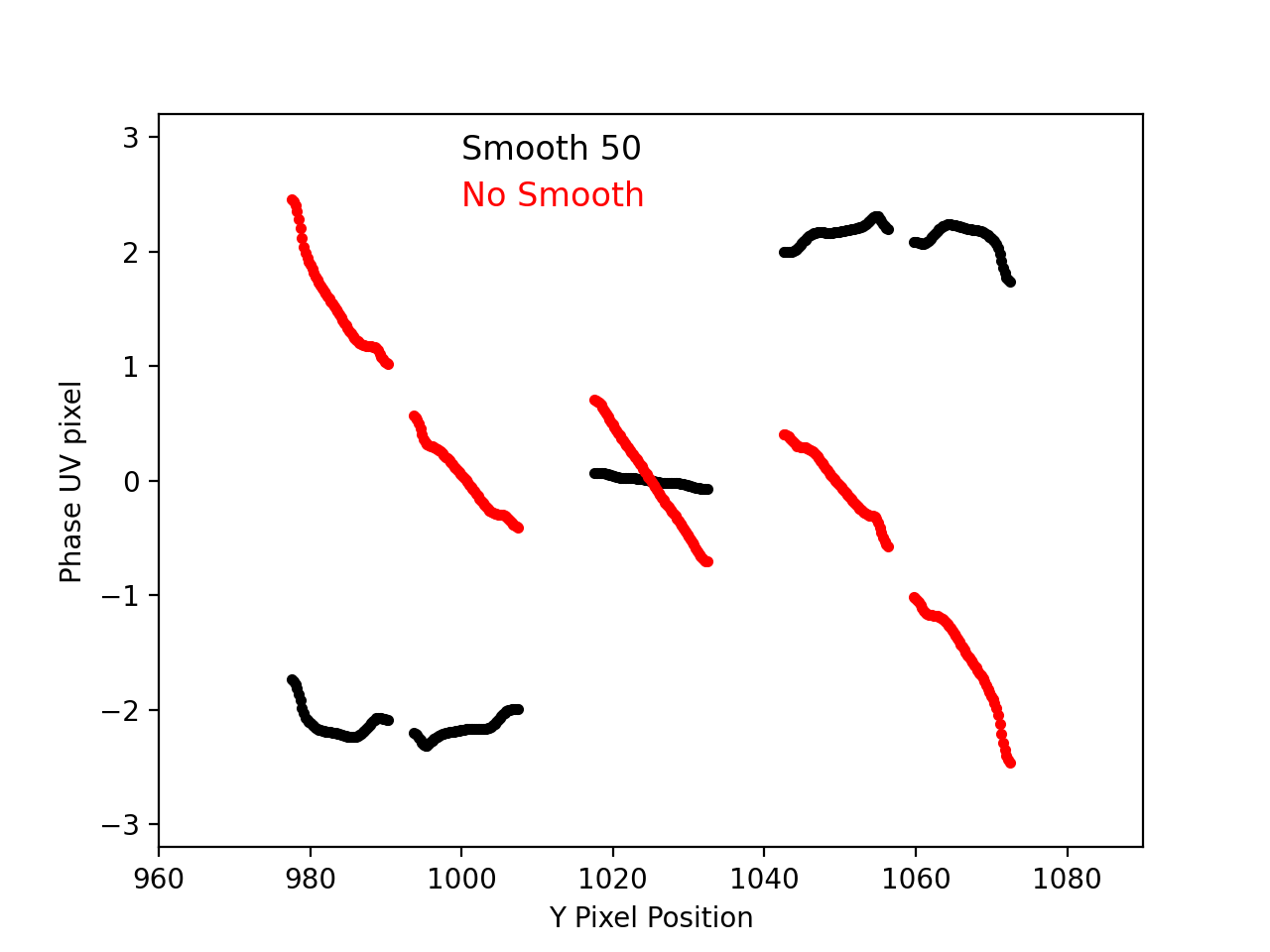}}
\caption{Cuts in Y direction through the u,v phase image, passing through the image center. Red is with peak image pixel centering, and black is with Airy disk centering. Each sub-segment corresponds to one of the u,v data points. Y-axis units are radians. 
}
\label{fig:Yphcut}
\end{figure}

Figure~\ref{fig:Yphcut} shows a cut in the Y direction across the phase distribution for different centering. The phase slopes are clearly reduced with centering on the Airy disk. 

Closure phase could be affected by centering of the image on the CCD -- the outer parts of the Airy disk, beyond the first null are sampled differently. For reference, the counts beyond the first null without bias subtraction contribute about 40\% to 45\% to the total counts in the field with 3\,mm holes and 1ms integations hole data. 

\begin{figure}[!htb]
\centering 
\centerline{\includegraphics[scale=0.5]{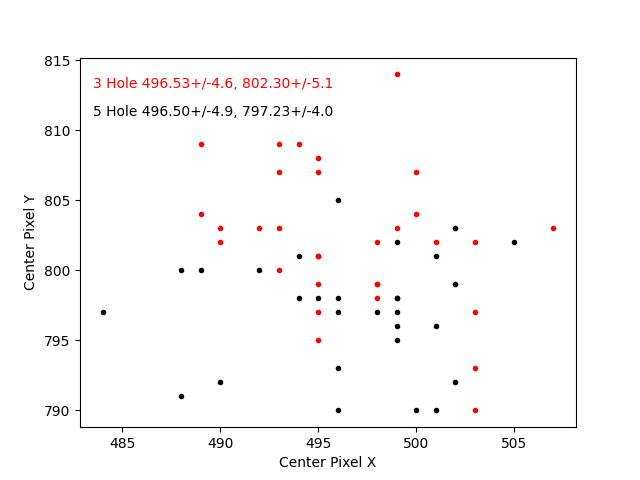}}
\caption{Image centering values for the 3-hole and 5-hole data with Airy disk centering
}
\label{fig:3H5Hcent}
\end{figure}

Figure~\ref{fig:3H5Hcent} shows the center pixel locations derived using Airy disk centering for the 3-hole and 5-hole data. The X values are the same. But the Y values differ by 5 pixels. The largest departures from zero closure phase for the 5-hole data all involve baseline 0-2, which is the 16\,mm vertical baseline (X direction in edf file which implies a narrow fringe in Y direction). This baseline is also in the 3-hole data, and it is the baseline with fringe length oriented horizontally, which might lead to the largest deviation in the case of a change in north-south centering of the fringe pattern on the CCD. 

\begin{figure}[!htb]
\centering 
\centerline{\includegraphics[scale=0.3]{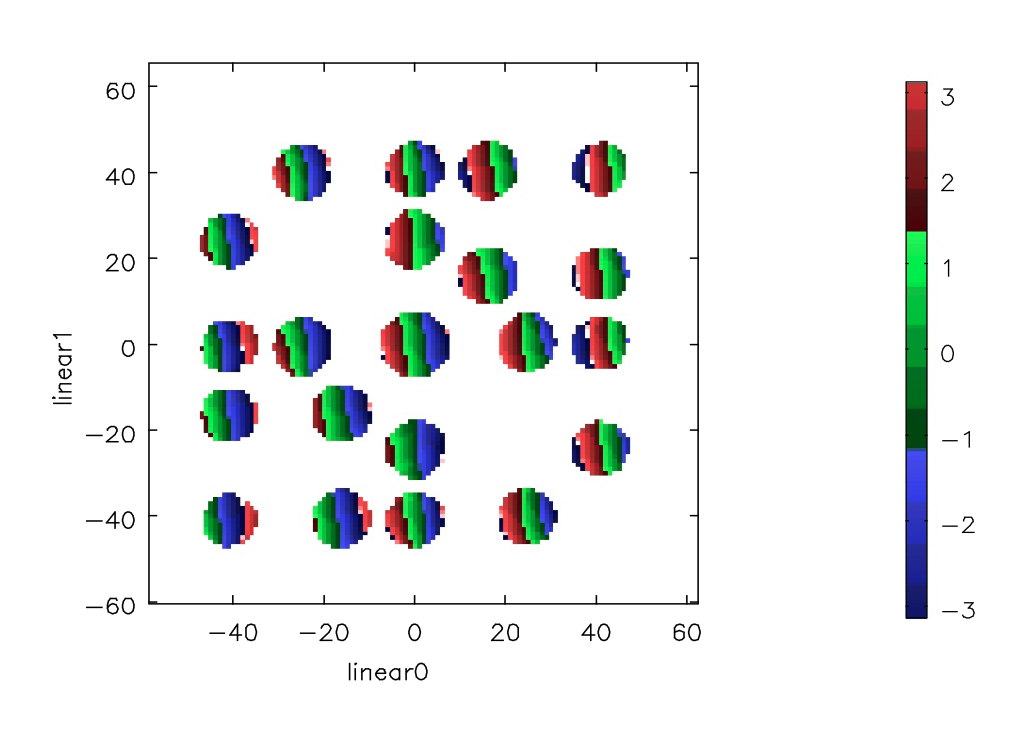}}
\caption{Image of 5-hole phase data without any image centering showing the complete phase wraps across the uv-aperture without centering (compare to Figure~\ref{fig:3mmfft}). Units are radians. 
}
\label{fig:WRAP}
\end{figure}

Figure~\ref{fig:WRAP} shows the resulting phase image without any centering. The offset of the image center from the CCD field center leads to a complete phase wrap across the uv-apertures. This compares to Figure~\ref{fig:3mmfft}, where only small phase gradients are seen after centering.

\subsection{U,V aperture radius: 3-9pix coherence and closure phases}
\label{sec:Npix}

\begin{figure}[!htb]
\centering 
\centerline{\includegraphics[scale=0.6]{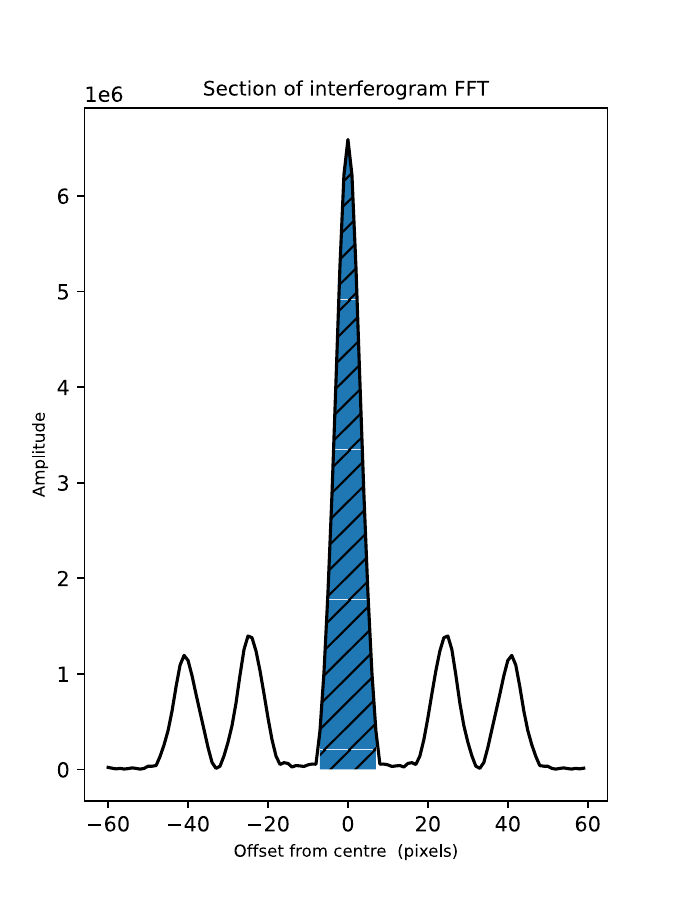}}
\caption{Section through the amplitude of the FFT of the example
  interferogram in Figure~\ref{fig:3mminterf}.  Hatched area
  illustrates the 7-pixel radius region used for estimating the
  visibilities.}
  \label{fig:3mmsection}
\end{figure}

We consider the radius of the size of the aperture in the u,v plane used to derive the amplitudes and phases of the visibilities. Figure~\ref{fig:3mmsection} shows a cut throught the center of the amplitude distribution of the u,v image. The hatched area shows the 7-pixel radius. This radius goes down to the 6\% point of the 'uv-beam'. Averaging beyond 9 pixels just adds noise, and beyond 10 pixel radius gets overlap between uv-measurements eg. 2-3 and 0-1.

\begin{figure}[!htb]
\centering 
\centerline{\includegraphics[scale=0.25]{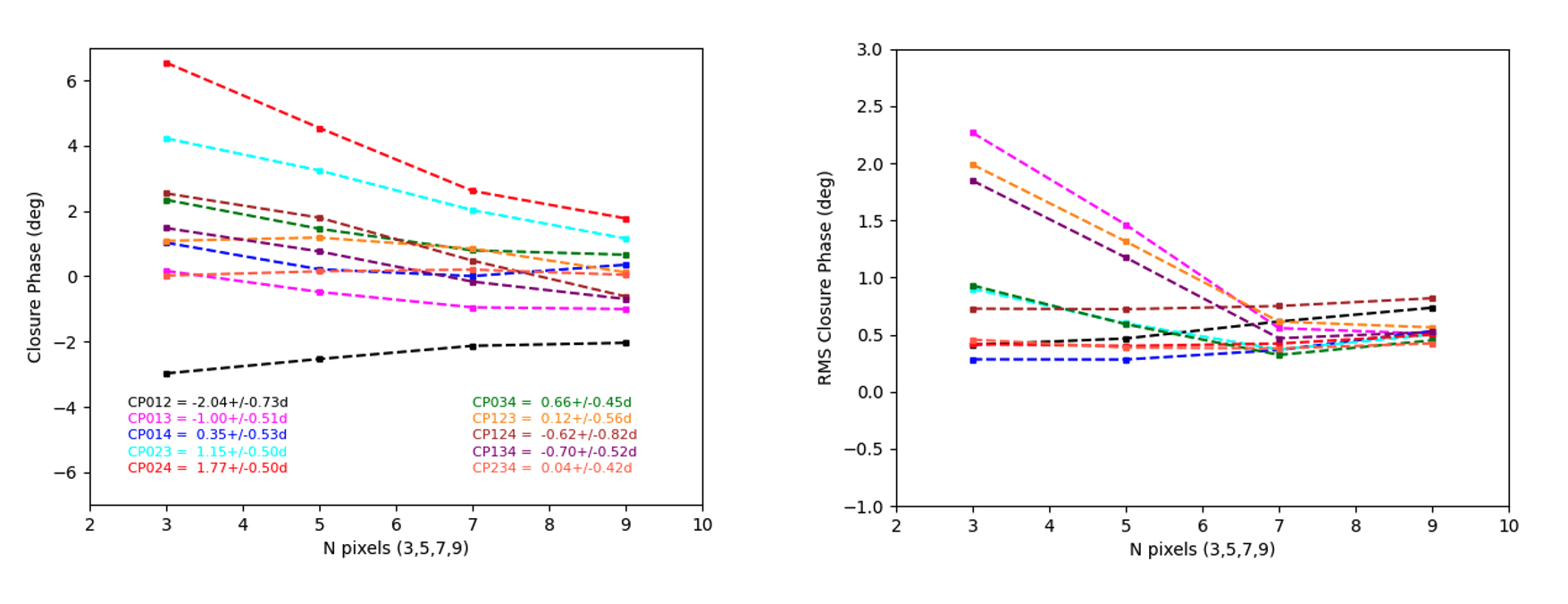}}
\caption{Left is the mean closure phase vs. N pixel radius of the uv-aperture for 3\,mm holes 1ms averaging, for the 10 Triads in the 5-hole mask. Right is the RMS scatter of the 30 records. }
\label{fig:CPvsNpix}
\end{figure}

We explore radii of 3, 5, 7, and 9 pixels, considering coherences and closure phases. Figure~\ref{fig:CPvsNpix} shows the closure phases versus the u,v aperture radius. The closure phase values tend toward smaller values with increasing aperture size. The RMS scatter decreases substantially with aperture size until 7pix radius. 

\begin{figure}[!htb]
\centering 
\centerline{\includegraphics[scale=0.3]{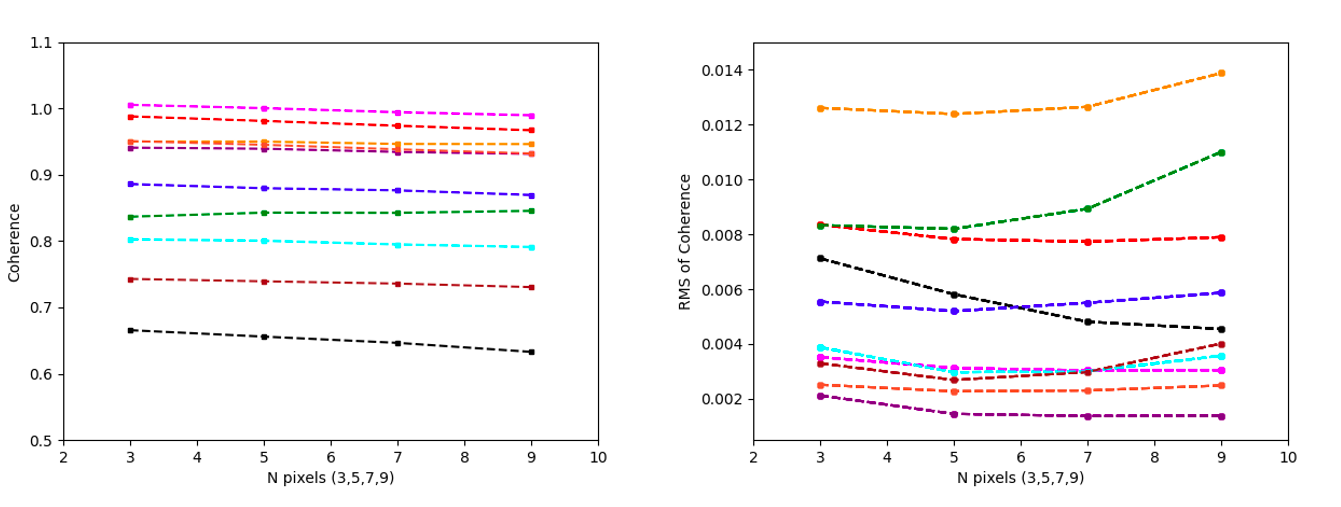}}
\caption{Left is the coherence vs. N pixel radius of the uv-aperture for 3\,mm holes 1ms averaging, for the 10 baselines in the 5-hole mask. Right is the RMS scatter of the 30 records. }
\label{fig:CohNpix2}
\end{figure}

Figure~\ref{fig:CohNpix2} shows the coherences for different u,v aperture radii. The coherences vary slightly, typically less than 2\%. The RMS of the coherences are relatively flat, or slightly declining, to 7 pixel radius, with a few then increasing at 9 pixels.

\subsection{3\,mm vs 5\,mm coherences}

\begin{figure}[!htb]
\centering 
\centerline{\includegraphics[scale=0.25]{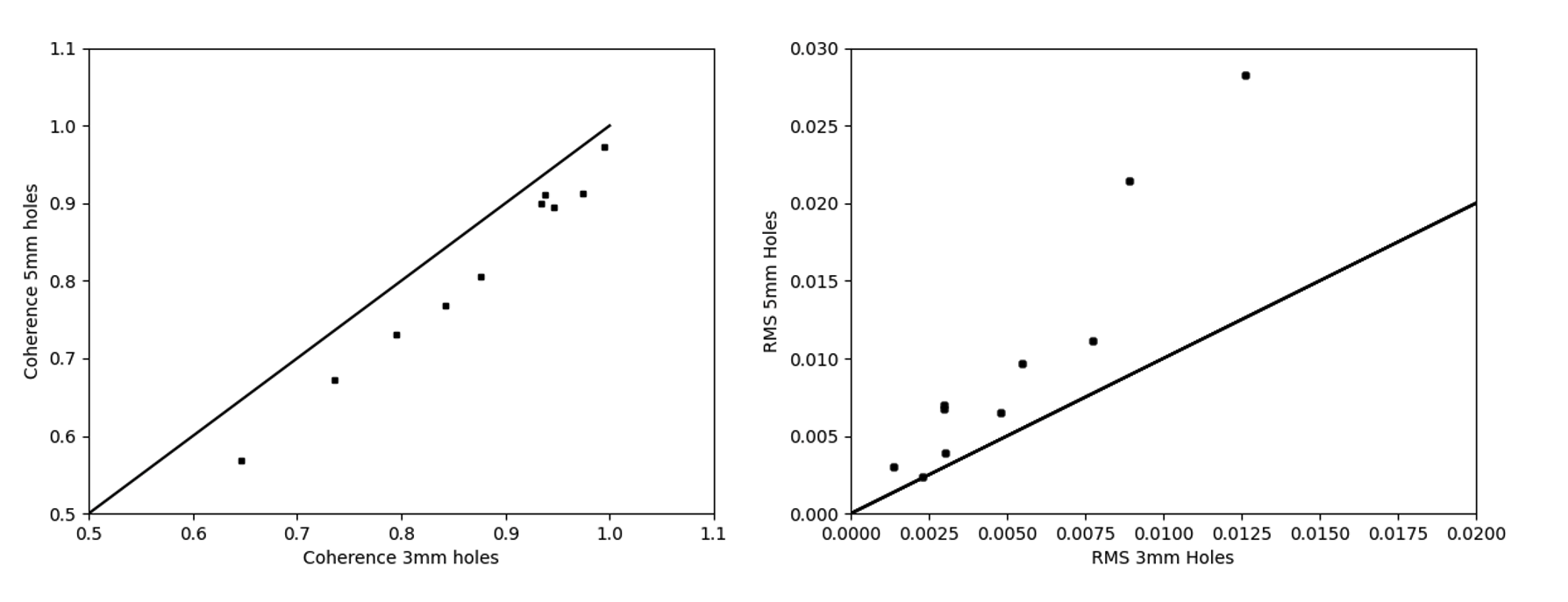}}
\caption{Left: A plot of measured coherence using 3\,mm holes vs 5\,mm holes and 1ms averaging for the 10 baselines in the 5-hole mask. The solid line would be equal coherence. Right: same but for the RMS of the coherence time series.}
\label{fig:cohhole}
\end{figure}

We consider the affect of the size of the hole in the non-redundant mask on coherence and closure phase. Figure~\ref{fig:cohhole} shows the coherence for a 5-hole mask with 3\,mm and 5\,mm holes. The 5\,mm data fall consistently below the equal coherence line, implying lower coherence by typically 5\% to 10\%. Also shown is the RMS for the coherence time series. The RMS scatter for the 5\,mm holes is higher, more than a factor two higher in some cases.

Lower coherence for larger holes may indicate phase gradients across  holes. A hole phase gradient is like a pointing error which implies mismatched primary beams in the image plane and may lead to decoherence.  

\subsection{3ms vs 1ms coherences: 5 hole data}
\label{sec:tave}

\begin{figure}[!htb]
\centering 
\centerline{\includegraphics[scale=0.35]{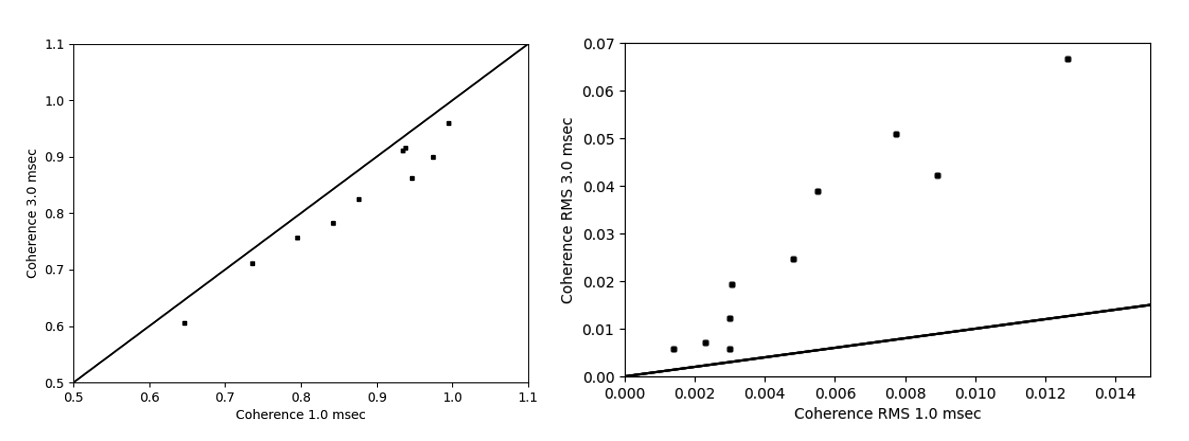}}
\caption{Left: 3\,ms vs 1\,ms coherence for the 5-hole mask. The solid line is equal coherence. Right: rms of the coherence for  3\,ms vs 1\,ms integrations.
}
\label{fig:cohtime}
\end{figure}

We consider the affect of the integration time on coherence and closure phase on the 5-hole data (see Section~\ref{sec:2hole} for further analysis with other masks). Figure~\ref{fig:cohtime} shows the coherence at 3\,ms vs 1\,ms integrations. The 3\,ms coherences are lower by about 2 - 10\%. The rms of 3\,ms coherences are much higher by factors 2 to 7. 

\begin{figure}[!htb]
\centering 
\centerline{\includegraphics[scale=0.33]{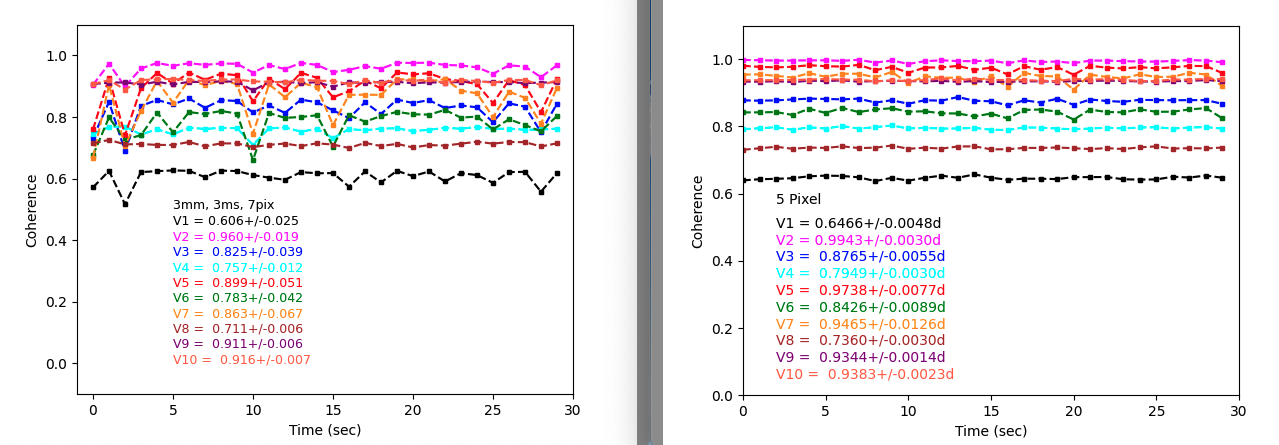}}
\caption{Plots of the time series of coherences for 5-hole data with 3\,ms and 1\,ms integrations.
}
\label{fig:explanation}
\end{figure}

The explanation of the Figure~\ref{fig:cohtime} is Figure~\ref{fig:explanation}, which shows the time series of coherences for 3\,ms vs 1\,ms. Two things occur: (i) the coherence goes down by up to 8\%, and (ii) the rms goes way up with 3\,ms, by up to a factor 7. The increased rms in 3\,ms data appears to be due to 'dropouts', or records when the coherence drops by up to 20\%. 

\begin{figure}[!htb]
\centering 
\centerline{\includegraphics[scale=0.25]{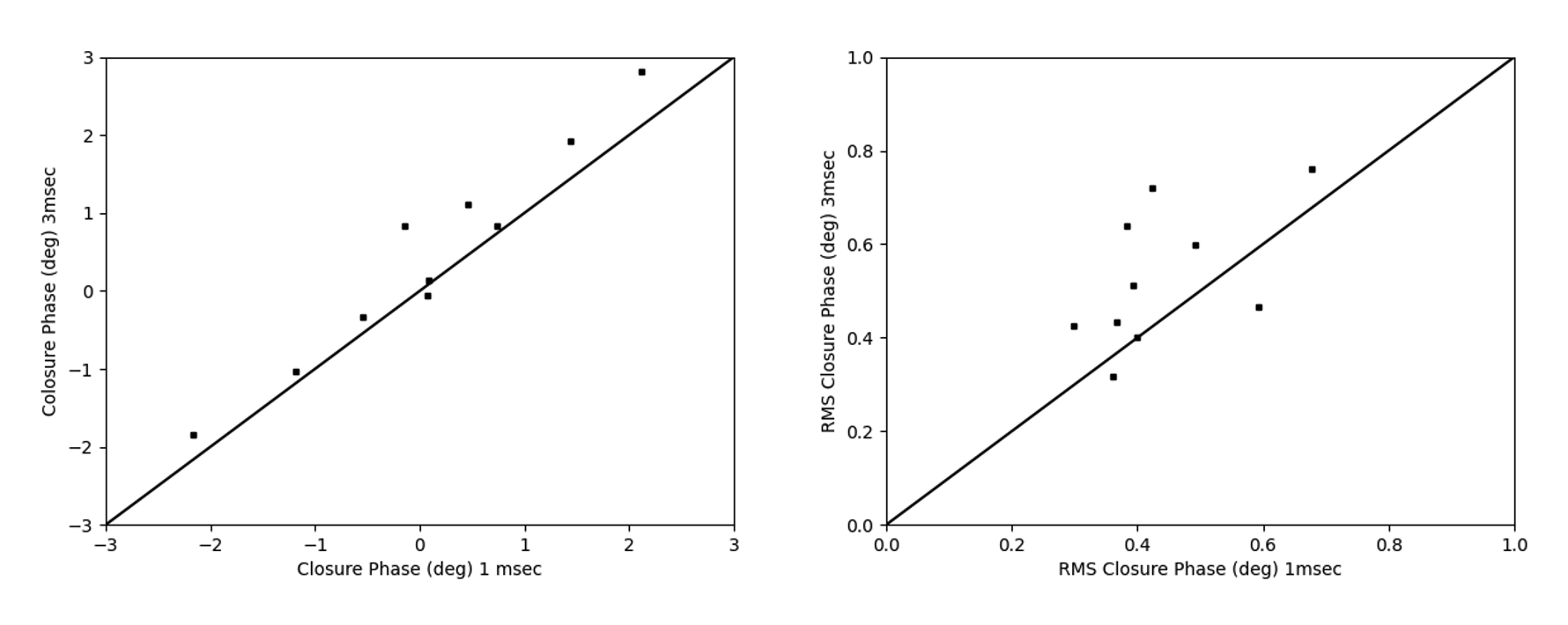}}
\caption{Left: 3\,ms vs 1\,ms closure phase for the 5-hole mask. The solid line is equal closure phase. Right: rms of the closure phase for  3\,ms vs 1\,ms integrations.
}
\label{fig:CP3msv1ms}
\end{figure}

Figure~\ref{fig:CP3msv1ms} shows the closure phases for 3\,ms averaging vs. 1\,ms averaging. The differences in closure phases are small, within a fraction of a degree. The rms scatter is slightly larger for 3\,ms, but again, not dramatically. Hence, closure phase seems to be more robust to averaging time, than coherence itself. 

\begin{figure}[!htb]
\centering 
\centerline{\includegraphics[scale=0.5]{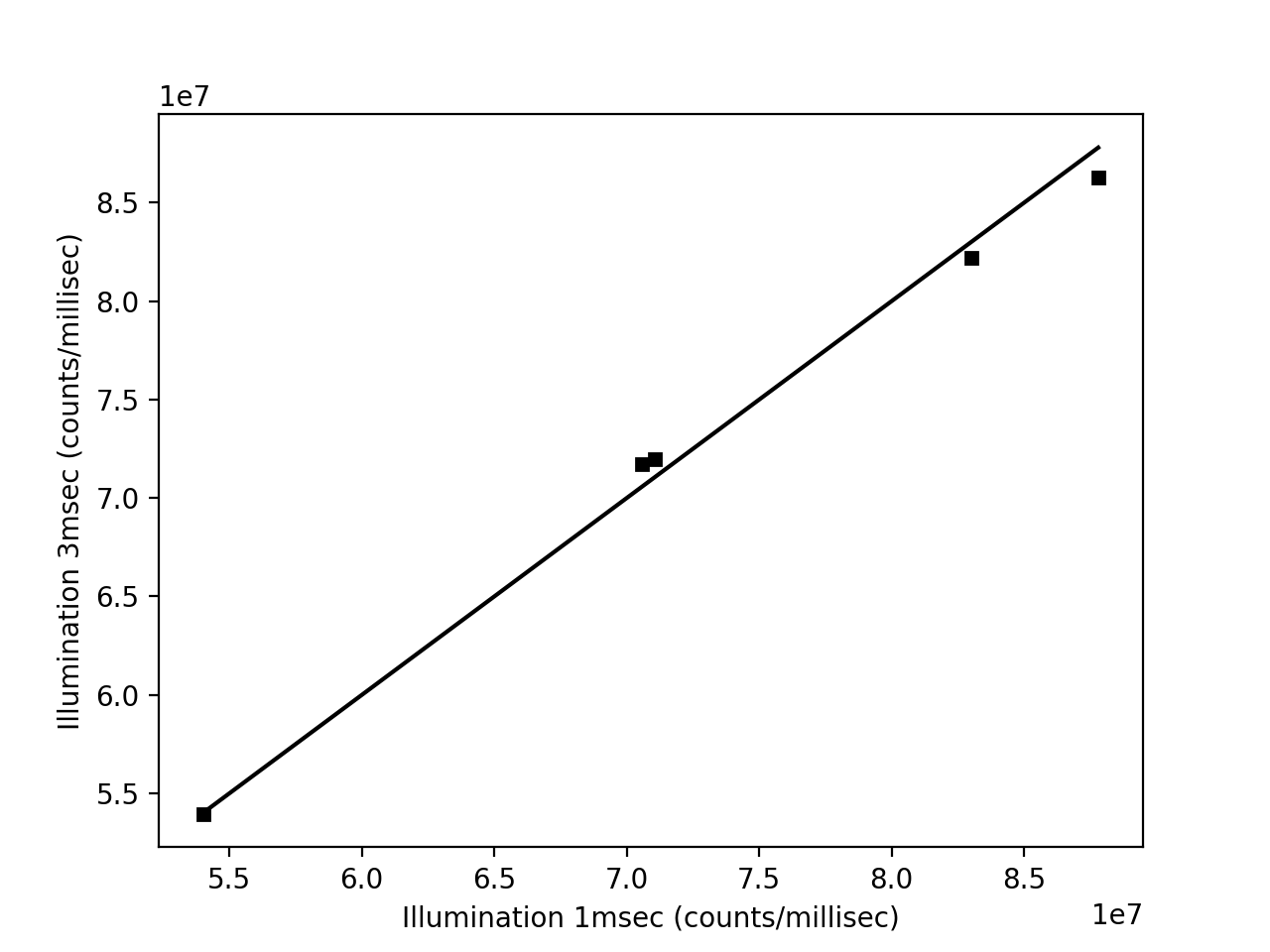}}
\caption{The illumination values for the 5-hole mask derived from the joint fitting procedure for 1\,ms and 3\,ms measurements. The line is equal illumination.
}
\label{fig:gains3ms1ms}
\end{figure}

To check if some of the decoherence of 3\,ms vs 1\,ms data could be caused by changing gain solutions in the joint fitting process, in Figure~\ref{fig:gains3ms1ms} we show the illumination values for the 5-holes derived from 1\,ms vs. 3\,ms data from the source fitting procedure. The illumination is defined as Gain$^2$, which converts the voltage gain from the fitting procedure into photon counts (ie. power vs. voltage). We then divide the 3\,ms counts by 3, for a comparison to 1ms data (ie. counts/millisecond).  Figure~\ref{fig:gains3ms1ms} shows that the derived illuminations are the same to within 2\%, at worst, which would not explain the 5\% to 10\% larger coherences for 1!ms data.

In Section~\ref{sec:2hole} we consider the effect of averaging time on all the data, including 2-hole and 3-hole measurements. 

\subsection{Bias subtraction}

We have calculated the off-source mean counts and rms for data using 2, 3, and 5-hole data, and for 1\~ms to 3\,ms averaging, and for 3~mm and 5~mm holes. The off-source mean ranged from 3.43 to 3.97 counts per pixel, with an rms scatter of 5 counts in all cases. We have adopted the mean value of 3.7 counts per pixel for the bias for all analyses. The bias appears to be independent of hole size, number of holes, and integration time, suggesting that the bias is dominated by a phenomenon such as CCD read noise.

\begin{figure}[!htb]
\centering 
\centerline{\includegraphics[scale=0.25]{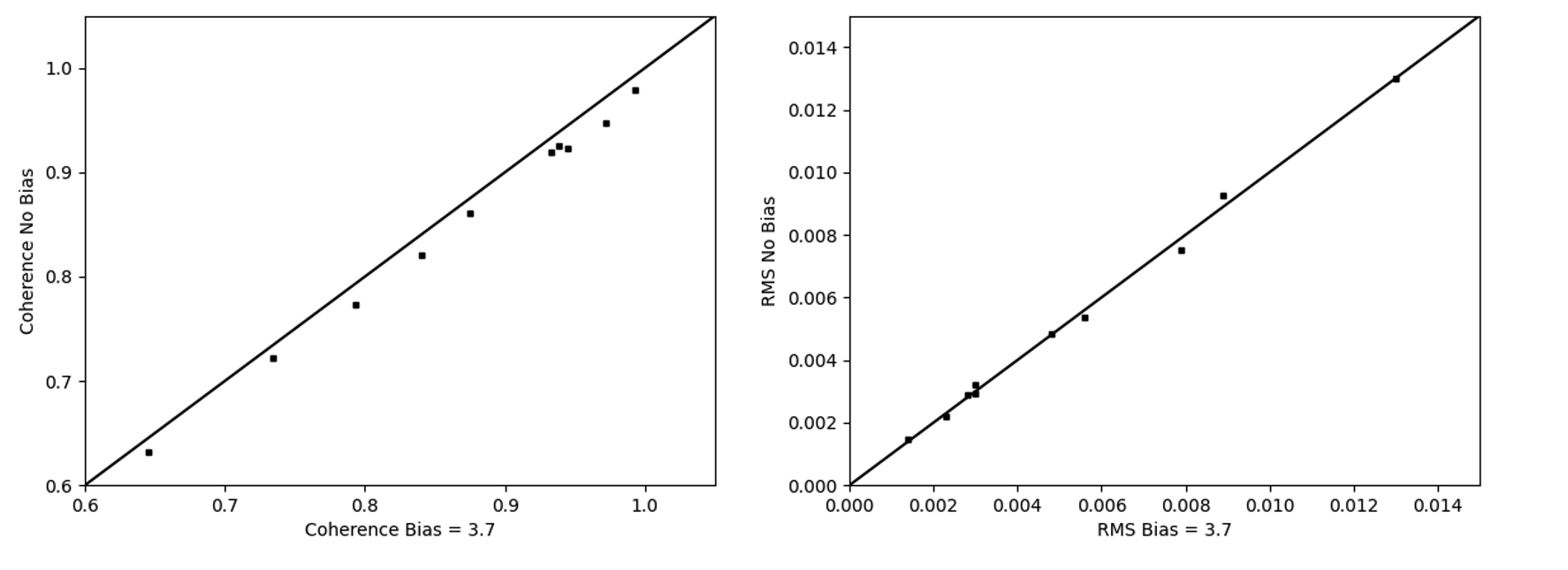}}
\caption{The mean coherences and rms of the coherences over the 30 record time series assuming no bias, and subtracting a bias of 3.7 counts per pixel. The solid line would be equal measurements. 
}
\label{fig:Cohbias}
\end{figure}

Figure~\ref{fig:Cohbias} shows a plot of the mean coherences and rms of the coherences over the 30 record time series assuming no bias, and subtracting a bias of 3.7 counts per pixel. The coherences are systematically lower by about 2\% with no bias subtracted, and the rms values are unchanged. 

\begin{figure}[!htb]
\centering 
\centerline{\includegraphics[scale=0.25]{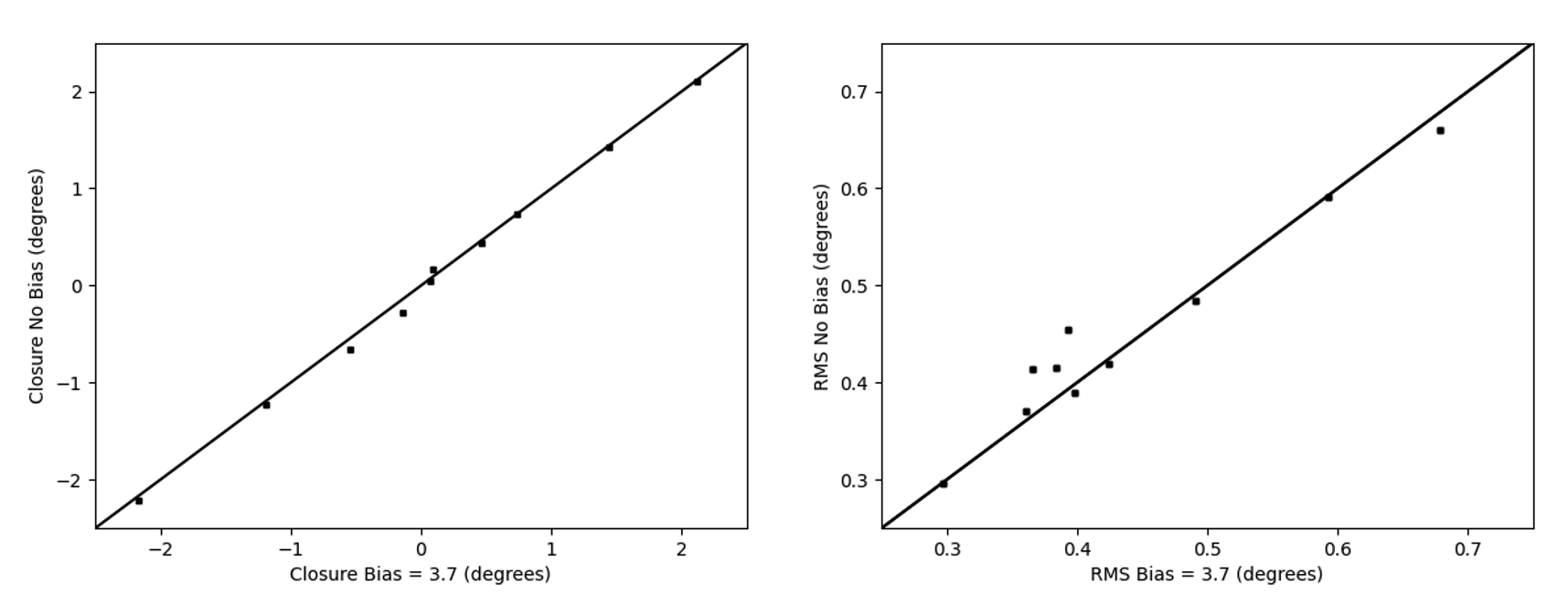}}
\caption{The mean closure phase and rms over the 30 record time series assuming no bias, and subtracting a bias of 3.7 counts per pixel. The solid line would be equal measurements. 
}
\label{fig:CPbias}
\end{figure}

Figure~\ref{fig:CPbias} shows a  plot of the closure phases over the 30 record time series assuming no bias, and subtracting a bias of 3.7 counts per pixel. The closure phases and scatter are effectively unchanged.

Overall, bias subtraction has a measurable, but not major, effect on the results of the interferometric measurements. Fortunately, the bias is directly measurable on the images, and hence we do not feel the bias is a significant source of uncertainty in our source size estimates.

\section{Decoherence due to Redundancy}
\label{sec:redundancy}

We demonstrate the decoherence caused by redundantly sampled visibilities using the 6-hole data. Recall that the 6-hole data has an outer square of holes, leading to two redundant baselines: the horizontal and vertical 16\,mm baselines, corresponding to [0-1 + 2-5] and [0-2 + 1-5]. In the Fourier domain, these redundant baselines sample the same spatial frequency. With no phase fluctuations, these two fringes will add in phase and roughly double the visibility amplitude (modulo the gain factors). But if there is a phase difference between the two fringes, then fringe contrast, or visibility coherence, will be lower (see Figure~\ref{fig:Fringesum}). 

The challenge is that one cannot independently determine the gain of hole 5 from the measurements, as was done for the five hole non-redundant data, since one cannot separate the gain factor and source size from the decoherence due to redundancy. As a start to the analysis, we investigate the time variability of the visibility amplitudes. We expect the variability of the redundantly sampled baselines should be higher than for non-redundant baselines given phase fluctuations and implied decoherence.

\begin{figure}[!htb]
\centering 
\centerline{\includegraphics[scale=0.6]{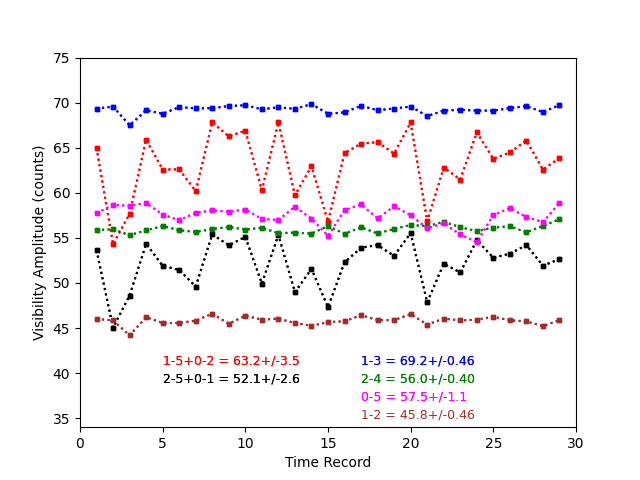}}
\caption{Visibility amplitudes for the time series for the 6-hole interferometric mask. Shown are both redundant (red, black) and non-redundant baselines. The redundant baseline amplitudes have been normalized by a factor two (consideration of coherences and gains is shown in Figure~\ref{fig:decoh}). The mean and rms scatter of the amplitudes are listed. 
}
\label{fig:6Hamps}
\end{figure}

Figure~\ref{fig:6Hamps} shows the results for a few of the visibility baselines for the 6-hole data. Shown are the two redundantly sampled baselines ([0-1 + 2-5] and [0-2 + 1-5]), and two non-redundant baselines that are similar in length and orientation to these redundant baselines (2-4 and 1-3, respectively). Also shown are the two longest baselines (rising and falling diagonals 1-2 and 0-5; see Figure~\ref{fig:mask6H}). We note that the visibility amplitudes for the non-redundant baselines in the 6-hole data are typically within 1\% of the same visibilities measured with the 5-hole mask. 

\begin{figure}[!htb]
\centering 
\centerline{\includegraphics[scale=0.6]{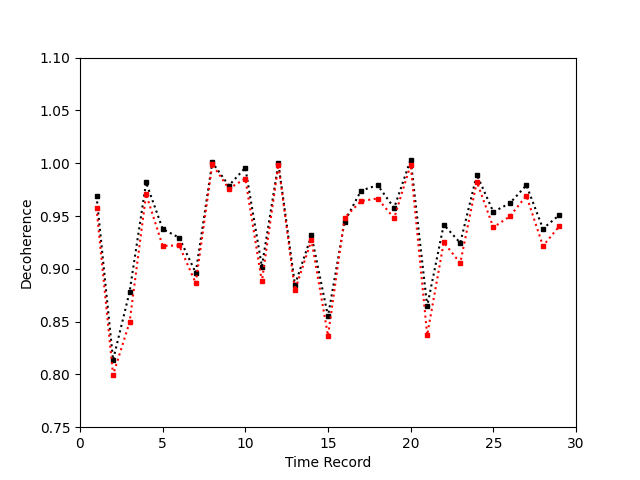}}
\caption{The decoherence of the redundantly sampled baselines (Black = [0-1 + 2-5] and Red = [0-2 + 1-5]), as calculated assuming a gain for hole 5 as described in Section~\ref{sec:redundancy}. 
}
\label{fig:decoh}
\end{figure}

A number of features are clear in Figure~\ref{fig:6Hamps}. First, the stability of the redundant baselines is much worse. The rms fluctuations with time are a factor 3 to 6 larger than for the non-redundant baselines. Second, the amplitudes of the non-redundant baselines that are similar in length and orientation to the redundant baselines are lower than for non-redundant baselines (red vs. blue and black vs. green; although better treatment of decoherence including the gains is given in Figure~\ref{fig:decoh}). Third, the time variations for the two redundant samples (black and red) are correlated. All these phenomena are consistent with decoherence of the redundantly sampled baselines due to aperture-based phase fluctuations.

To investigate decoherence caused by redundant sampling, we assume a gain for hole 5 equal to the mean between the gains of the two holes in close proximity to hole 5 (holes 3 and 4 see Figure~\ref{fig:mask6H} and Table~\ref{fig:gains}, mean gain of 3 and 4 = G$_5$ = 9.25). This gain assumes that the illumination pattern in that corner of the mask is relatively uniform. In this case, the decoherence becomes the ratio of the measured visibility amplitude (V$_{\rm 6Hmeasured}$), to the sum of the expected visibility amplitudes of the redundant samples (ie. assuming no phase difference and decoherence). The expected amplitudes are given by: $\rm V_{i,j} = \gamma_{i,j} G_i G_j$, where the gains were derived from the 5-hole non-redundant mask (Table~\ref{fig:gains}; and assuming G$_5$ = 9.25), and the coherence is also measured for the given baseline using the 5-hole non-redundant mask (see  Table~\ref{fig:cohtab}). For example, for redundant sample [0-1 + 2-5], the decoherence is: 

$$\rm Decoherence = V_{6Hmeasured}/(\gamma_{0,1}G_0 G_1  + \gamma_{0,1} G_2 G_5) $$

Figure~\ref{fig:decoh} shows the decoherence time series for the two redundantly sampled visibilities. Again, the scatter is substantial, as seen in Figure~\ref{fig:6Hamps}. The mean and rms values for visibility [0-2 + 1-5] $= 0.931\pm 0.052$, while those for [0-1 + 2-5] are $0.942\pm 0.047$. Note that the maximum decoherence ratio reaches a value of unity, as expected for no phase decoherence, ie. when the two redundant visibilities are in-phase. This maximum of unity lends some confidence in the assumed gain for hole 5.

One curious result is the close correlation between the decoherence of the two redundant baselines with time, as can be seen in Figure~\ref{fig:decoh} and Figure~\ref{fig:6Hamps}. Some correlation is expected, since the phase fluctations at hole 5 are common to both baselines. But we are surprised by the degree of correlation. Perhaps vibrations of optical components might be more susceptible to such close correlation as opposed to laboratory 'seeing'? Further experiments are required to understand the origin of visibility phase fluctuations in the SRI measurements. 

We conclude that redundant sampling of the visibilities leads to decoherence at the level $\sim 5\%$, with a comparable magnitude for the scatter for the time series. A 5\% decoherence is comparable to that seen comparing 1\,ms vs. 3\,ms time averaging of interferograms (Figure~\ref{fig:cohtime}), and likewise the larger rms scatter of the time series is similar to that seen comparing 1\,ms and 3\,ms averaging. A 5\% decoherence for a redundantly sampled baseline would be caused by a $\sim 20^o$ phase difference between the two redundant visibilities.

These results necessitate the use of a non-redundant mask to avoid decoherence caused by hole-based phase perturbations due to eg. turbulence in the lab atmosphere or vibration of optical components (Torino \& Iriso 2015). Likewise, a filled-aperture imaging system will display image smearing due to these phase perturbations. 

\section{Time Averaging in 2-hole, 3-hole, and 5-hole Measurements}
\label{sec:2hole}

A curious result from the source size analysis (Nikolic et al. 2024), was that the electron beam size derived with from the 5-hole mask with 3~ms averaging data resulted in a larger derived beam size than the 5-hole 1~ms data. We initially assumed this was due to temporal decoherence of the 3~ms data. However, the rotating 2-hole technique, using 3~ms averaging, resulted in the correct beam size, although with larger scatter in the time series. We explore this apparent contradiction using the 3-hole data as a third comparator. 

\begin{figure}[!htb]
\centering 
\centerline{\includegraphics[scale=0.3]{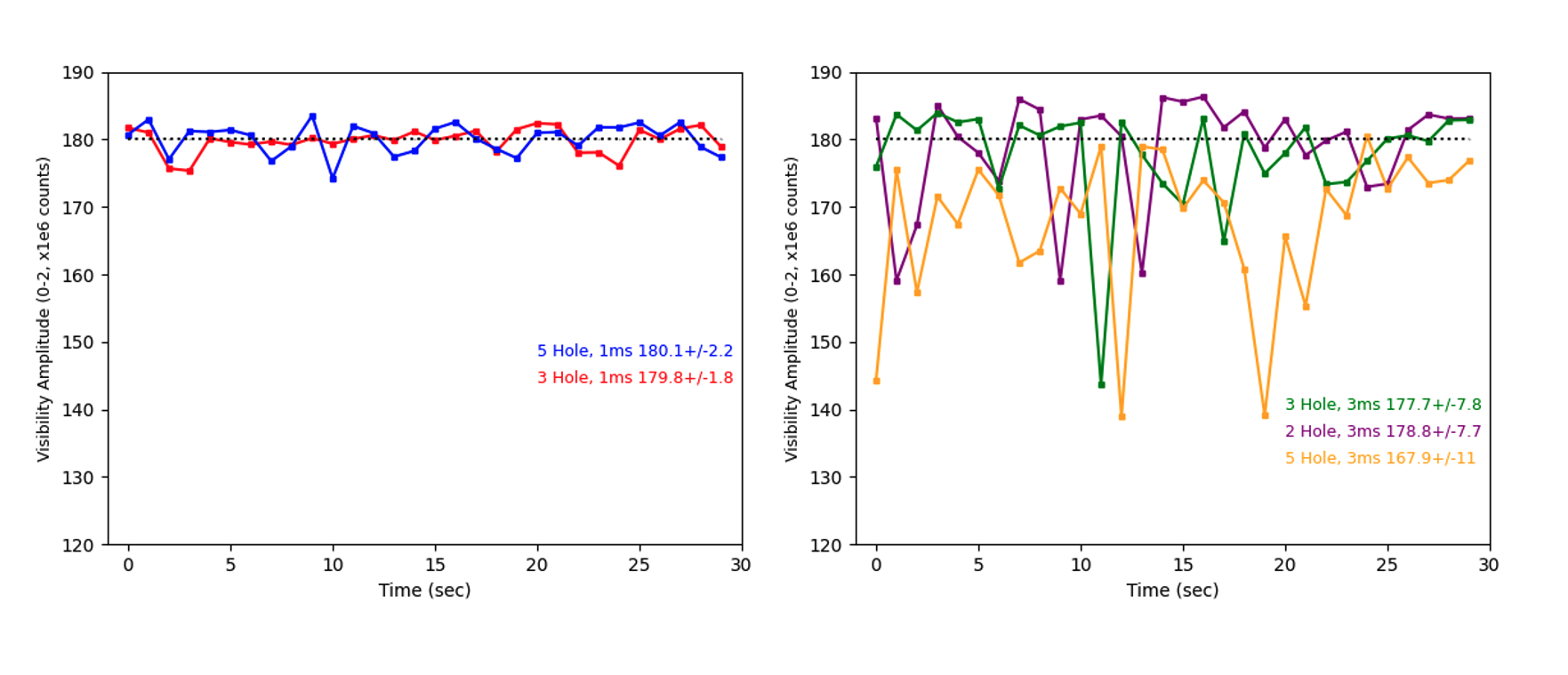}}
\caption{The measured visibility amplitude, in counts, for the time series of measurements, for the 16\,mm vertical baseline (0-2). Left: 1\,ms averaging data for the 5-hole and 3-hole masks with 3\,mm apertures. Amplitudes for the 1\,ms data have been multiplied by a factor 3 to normalize for integration time relative to the 3\,ms data. Right: 3\,ms averaging data for 2-hole, 3-hole, and 5-hole masks. The mean and values and color coding are listed on the plot. The Y-axis scales are the same, and the dotted line shows a mean value of 180e6 counts, for reference. 
}
\label{fig:Visamps235}
\end{figure}

In Figure~\ref{fig:Visamps235} we show the time series of the measured visibility amplitudes (in counts) on the 16\,mm vertical baseline (0-2) for the following data: 5-hole 1\,ms and 3\,ms, 3-hole 1\,ms and 3\,ms, and the 2-hole 3\,ms (recall, no 2-hole 1~ms data was taken). This is a busy plot, but the main points can be summarized as: 

\begin{itemize}

\item The 3-hole 1ms and 5-hole 1ms visibility amplitudes (red and blue) agree nicely, with less than a percent difference in mean value of $180\pm 2$, with no dropouts and low scatter.

\item The 2-hole 3ms and 3-hole 3ms visibilities (green and purple) also have most points of similar amplitude as the 5-hole 1ms data, with a mean value of $\sim 178\pm 8$, but the rms scatter is larger than the 1\,ms data by a factor four. This larger scatter is partially due to a few points appearing as 'dropouts', with amplitudes up to 20\% lower than the rest. 

\item The outlier dataset is the 5-hole 3ms data (yellow). This data set also has dropouts (see also Figure~\ref{fig:explanation}), but further, all of the points appear low, with a mean value substantially lower than all the other data sets, and with the largest rms scatter: $168\pm 11$.

\end{itemize}

Hence, comparing the 1ms data (5 and 3 holes), to the 3ms data (2 and 3 holes), it would appear that the primary effect of longer integration time is to increase the scatter, and cause a few substantial dropouts, which lowers the mean amplitude slightly. But most amplitudes are consistent between 1ms and 3ms frame times, leading to the consistency between the 2-hole 3ms source size calculation and the 5-hole 1ms result. 

The 5-hole 3ms data is anomalous, in that it has the highest scatter, including dropouts, and a 7\% lower overall mean. We speculate that the turbulence or other phase corrupting factors in the laboratory (eg. mirror vibrations), were different at the time of the 5-hole 3ms experiment. Further experiments are planned to explore this issue.

\begin{figure}[!htb]
\centering 
\centerline{\includegraphics[scale=0.2]{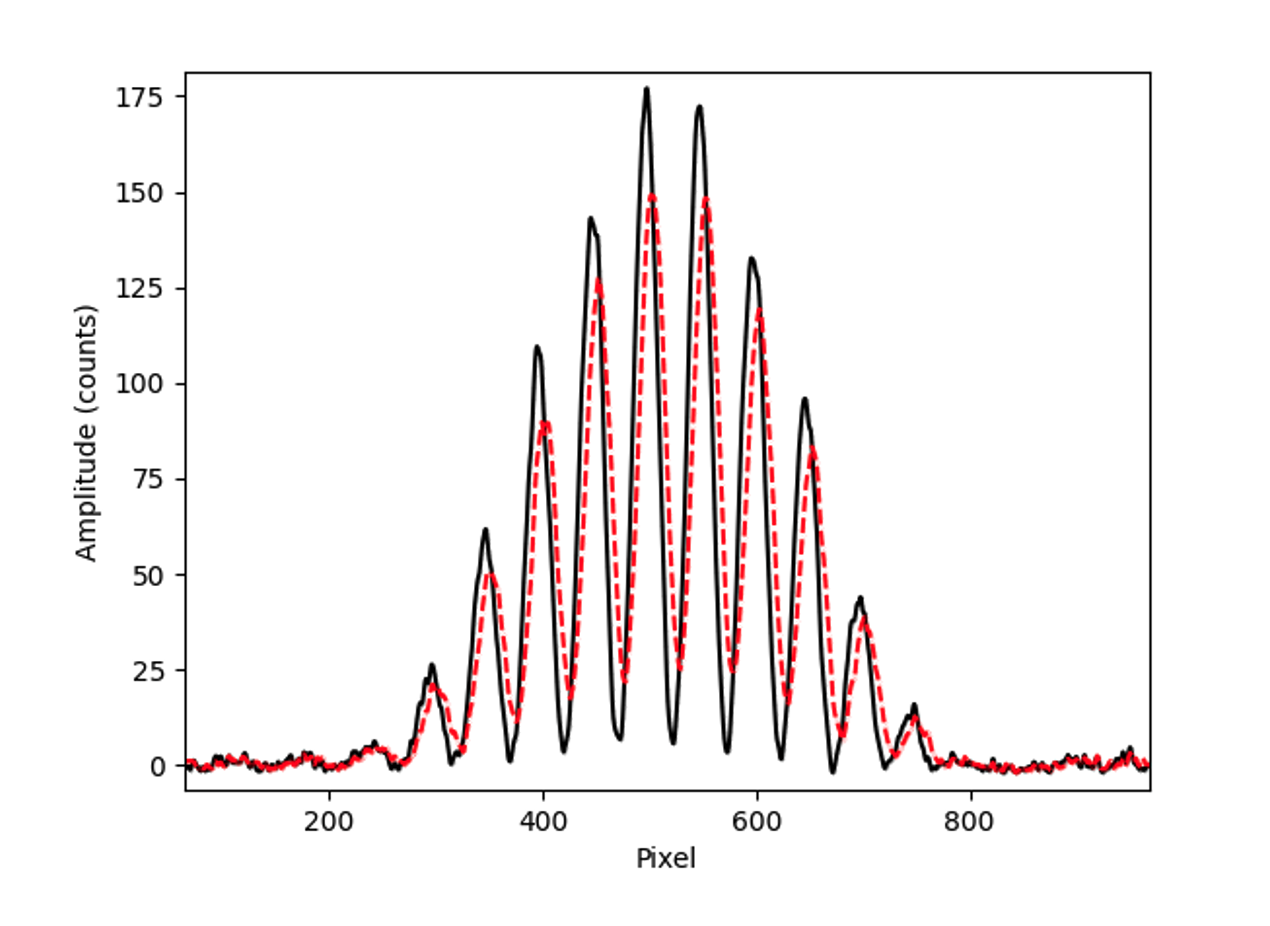}}
\caption{The black curve shows the fringe profile for the 2-hole vertical 16\,mm fringe. The red curve shows the profile after shifting the image by 1/4 wavelength, then summing with the original and scaling by a factor 1/2. 
}
\label{fig:Fringesum}
\end{figure}

As a final pedagogic exercise, we ask: what happens when, for a given baseline, two fringes of equal amplitude but different phase (= position on detector) are summed during a single measurement, as would occur, for instance, if during the frame integration time there is a phase perturbation leading to a rigid shift of the fringe. Figure~\ref{fig:Fringesum} shows the profile through a vertical fringe from the 2-hole data, as well as the same profile after shifting the frame by 1/4 of the fringe separation, then summing and scaling by a factor 1/2. The result is not a smearing or broadening of the fringe, nor a change in fringe spacing, but a change in the contrast, or the ratio of the maximum to minimum values. This behaviour is the basis for the historical definition of the fringe coherence, or visibility = $\rm (I_{max} - I_{min})/(I_{max} + I_{min})$ (Michelson 1890; Monnier 2003). 

\section{Error Analysis}
\label{sec:errors}

\subsection{Photon Noise Floor}

We have tens of millions of CCD counts per image, and hence the errors from photon counting statistics are low.  To obtain a rough estimate, we perform two calculations.

First, the number of photons contributing to a visibility is roughly the sum in
the uv-aperture divided by the number of pixels in the uv-aperture (the finite size of the uv-aperture is due to a convolution in the uv-plane by the Fourier transform of the 'primary beam', i.e. the Airy disk, and hence the uv-pixels are not independent). Typically, the visibility amplitudes integrated over the 7-pixel uv-aperture radius, are order $6\times 10^7$, implying about $3.8\times 10^5$ counts. The fractional error from photon counting  is then 1/root($3.8\times 10^5$) = 0.16\% .

Second, we sum over similar 7-pixel radius apertures in regions of the uv-plane with no signal, and get a mean value of $\sim 1.5\times 10^5$, which, relative to the typical visibility-aperture values of $6\times 10^7$ implies an error of $\sim 0.25\%$

\subsection{Processing Errors: Gaussian Random Approximation}

Beyond photon statistics, there are a number of processing steps that affect the resulting coherences, and hence the fit to the source size, including: uv-aperture size, bias subtraction, image centering, and others. In this section, we perform modeling of the uv-data to get an estimate of what level of errors in the coherences could lead to the measured scatter in the final results, assuming a Gaussian random distribution for the various errors over time. Systematic errors with time are considered below. While not strictly rigorous (the modeling does not include effects related to eg. the edges of the CCD or bias subtraction), this uv-model approach does provide a rough estimate of the summed level of error likely in the ALBA data, as well as how such errors may affect the final results.

We start by creating a FITS image of a Gaussian model with the shape of the ALBA electron beam, for which we adopt a dispersion of 60$\mu$m $\times$ $24\mu$m, and major axis position angle $= +16^o$ CCW from the horizontal. This model image is converted into arcseconds using the distance between the mask and the synchrotron source (15.05~m). We also generate a configuration file corresponding to the 5-hole mask used in our experiments, with baselines and hole size scaled to get uv-coordinates in wavelengths. A uv-data measurement set is then generated from the model and the configuration using the CASA task 'SIMOBSERVE', resulting in a 10 visibility measurement set with the proper uv-baseline distribution, primary beam size, and model visibilities (complex coherences). 

Gaussian random noise is then added to the complex visibilities at the rms level of $\sim 10\%$ of the visibility amplitudes, and a second test was done with $1\%$ rms noise. Since the noise is incorporated in the complex visibilities, it affects both phase and amplitude. In each case, a series of 30 measurement sets with independent noise (changing 'setseed' parameter), are generated to imitate the 30 frames taken in our measured time series. 

We employ 'UVMODELFIT' in CASA to then fit for the source amplitude, major axis, minor axis, and major axis position angle. Starting guesses are given that are close to, but not identical with, the model parameters (within 20\%), although the results are insensitive to the starting guesses (within reason). 

We first run uvmodelfit on the data with no noise, and recover the expected model parameters to better than $10^{-3}$ precision. These low level differences arise from numerical pixelization. 

\begin{figure}[!htb]
\centering 
\centerline{\includegraphics[scale=0.32]{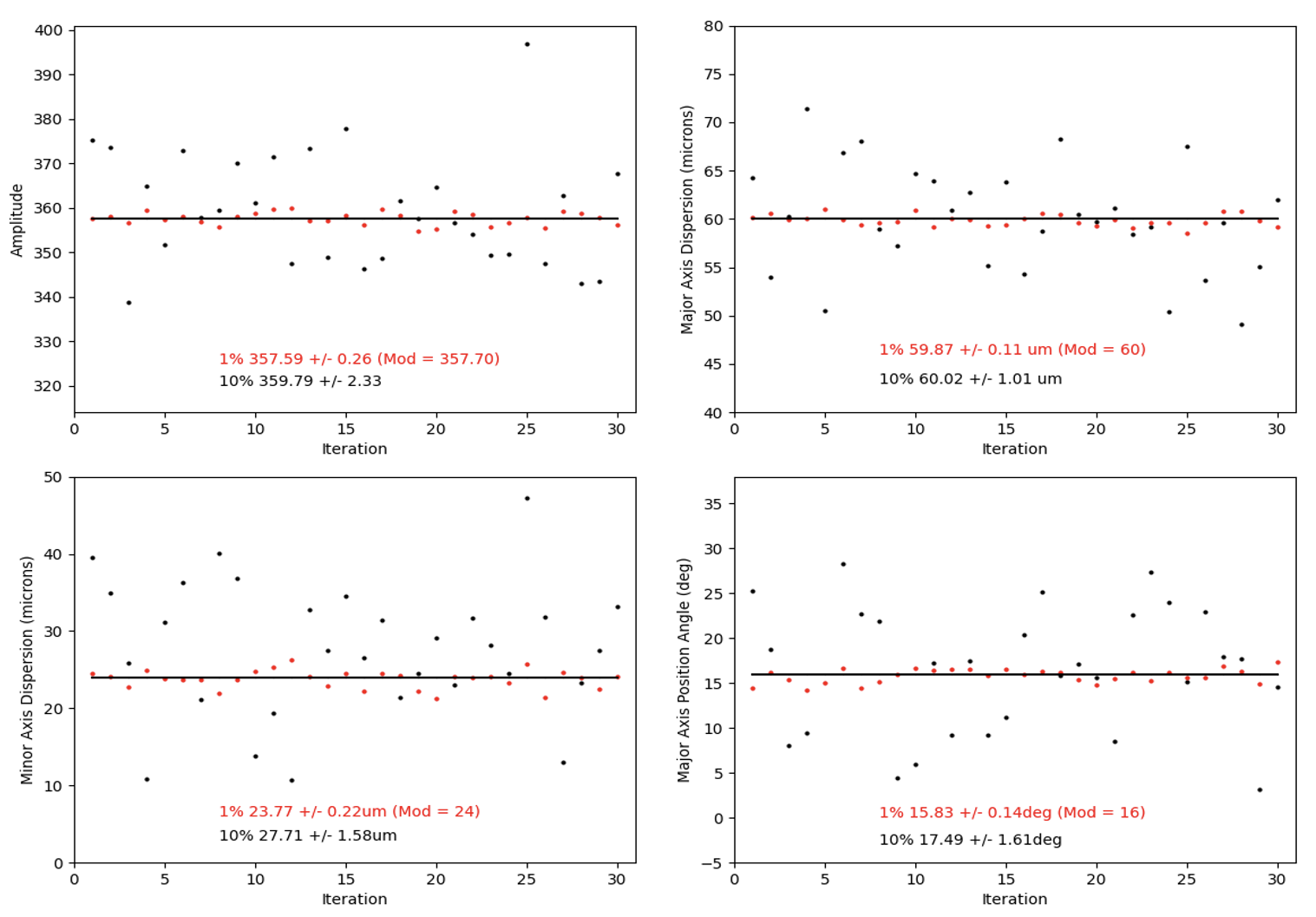}}
\caption{Results from modeling of the uv-data with errors. The black points are for 10\% Gaussian random errors on the complex visibilities, and the red points are 1\% errors. The solid line is the input model.  
}
\label{fig:error}
\end{figure}

Figure~\ref{fig:error} shows the results for the two simulation 'time series', and Table~\ref{fig:errortab} lists the values for the mean and rms/root(30). Also listed are the results from the measurements in Nikolic et al. (2024), and the input model. Two results are of note.

First, the 1\% rms noise on the visibilities results in fitted quantities (amplitude, bmaj, bmin, pa), that are consistent with the model parameters, to within the scatter. Also, the rms scatter for the fit paramaters are of similar magnitude as those found for the real data. 

Second, the 10\% rms visibility noise leads to $\sim 10\times$ higher scatter in the fitting results. Moreover, the minor axis fitting shows a skewed distribution toward values higher than the input model (ie. 21 points above the model line, and 9 below). We suspect that this skewness arises due to a Poisson-like bias when fitting for a positive definite quantity, when the errors become significant compared to the value itself. This skewness is not seen in the 1\% error analysis.

\begin{table}
  \centering 
\begin{tabular}{lcccc}
\toprule
~ & Amplitude & Major Axis & Minor Axis & Position Angle \\
~ & ~ & microns & microns & degrees \\
  \midrule
Data Fit & ~ & ~~$59.6\pm 0.1$~~ & ~~$23.8\pm 0.5$~~ & ~~$15.9\pm 0.2$ \\
Model & 357.70 & 60 & 24 & 16 \\ 
1\% errors & ~~$357.59\pm 0.26$~~ & $59.87\pm 0.11$ & $23.77\pm 0.22$ & $15.83\pm 0.14$ \\
10\% errors & $359.79\pm 2.33$ & $60.02\pm 1.01$ & $27.71\pm 1.58$ & $17.49\pm 1.61$ \\
\bottomrule
\end{tabular}
\caption{Error analysis from modeling. The first row lists the measurements from Nikolic et al. (2024)}
\label{fig:errortab}
\end{table}

\subsection{Systematic Errors}

We investigate the effect of systematic errors using real data for visibility amplitudes from the ALBA 5-hole data. Two simple tests are performed: adjust the amplitudes systematically low by 5\%, and high by 5\%, then run the Gaussian fitting routine in Nikolic et al. (2024). 

The fitted source size for the 5\% low amplitudes increases by 6.4\%, while the size decreases by 6.9\% for the 5\% high amplitudes. This is qualitatively consistent with the expected increase in source size for lower coherences, and vice versa. Quantitatively, for small offsets, the source size appears to be roughly linear with systematic offset for the visibility amplitudes. However, the fitting routine includes a joint fit for the gains of each hole. These gains also change slightly with systematic errors, with up to 2\% lower gains for lower amplitudes, and similarly higher gains for higher amplitudes. 

\section{Summary and Future directions}

\subsection{Summary}

We have described processing and Fourier analysis of multi-hole interferometric imaging at optical wavelengths at the ALBA synchrotron light source to derive the size and shape of the electron beam using non-redundant masks of 2, 3, and 5 holes, plus a 6-hole mask with some redundancy. The techniques employed parallel those used in astronomical interferometry, with the addition of gain amplitude self-calibration. Self-calibration is possible in the laboratory case due to the vastly higher number of photons available relative to the astronomical case. We have considered varying hole size and varying frame time. The main conclusions from this work are:

\begin{itemize}

\item The size of the Airy disk behaves as expected for changing hole sizes. There are many photons (millions), such that the diffraction pattern is sampled beyond the first null of the Airy disk, to the edge of the CCD field.

\item We develop a technique of self-calibration assuming a Gaussian model to simultaneously solve for the source size and the relative illumination of the mask (the hole-based voltage gains). The gains are stable to within 1\% over 30 seconds, and relative illumination of different holes can differ by up to 30\% in voltage solutions. Hence, gain corrections are required to derive  visibility coherences, and hence the source size.

\item We show visibility phases have a peak-to-peak variation over 30 seconds of $\sim 50^o$. Further, coherences for 3\,ms frame-times for the 5-hole data are systematically lower than those for 1\,ms frame time by up to 10\%, and the 3\,ms coherences are much noisier than 1\,ms. We also find the phase fluctuations are correlated on two longer and similar baselines.

\item We find the closure phases are remarkably stable, with an RMS $\sim 0.3^o$ over 30 seconds, even though the visibility phases show large temporal variations. This implies that the visibility phase variations can be factorized into phase terms associated with the complex voltage response of each hole in the mask. 

\item The closure phases are small, $\le 2^o$. For a well resolved source, small closure phases imply the source structure is symmetric to within $\sim 1\%$ of the total flux from the source. However, for a marginally resolved source, small closure phases may still arise regardless of complex source structure on scales much less that the fringe spacing. 

\item We demonstrate decoherence on redundantly sampled baselines using a 6-hole mask. The redundantly sampled baselines show decoherence at the level of $\sim 5\%$, and have a much larger scatter of the visibility amplitudes for the time series. The results are consistent with phase fluctuations affecting the coherence of redundantly sampled baselines. These fluctuations necessitate the use of  non-redundant masks, as opposed to redundant masks or filled apertures, in order to properly measure the coherences and source size. 

\item Comparing visibility amplitudes for the 0-2 baseline (16 mm vertical) 2-hole 3ms data with 3-hole 1ms and 3ms data, and with the 5-hole 1ms data, the primary effects of the longer integration are to induce 'dropouts', where the coherence falls by 5\% to 15\% for one 3ms frame, and to increase the overall rms scatter of the time series. The mean value of the amplitudes are consistent between the 1ms and 3ms 2-hole and 3-hole data, although the scatter is a factor 4 higher for the 3ms data. The one outlier data set is the 5-hole 3ms data, where dropouts are also seen, but the mean amplitude for the rest of the points is also lower by about 5\%, and the rms scatter is the largest by 40\%. 

\item We estimate a noise floor for the visibilities based on photon statistics of order 0.2\%. However, there are a number of steps in the processing that could lead to larger errors. We perform an error analysis based on model visibilities and Gaussian fitting which suggests that the noise on our measured visibilities is of order $1\%$. Investigation of larger errors shows a bias toward larger fitted minor axis values, likely due to a Poisson-type bias. We find the fitted source size is roughly linear with systematic offsets of the coherences, for small offsets. 

\end{itemize}

In terms of processing, we find:

\begin{itemize} 

\item Centering on the Airy disk of the measured images leads to the best results in the Fourier domain, meaning, the smallest phase slopes across the u,v aperture.
 
\item A 7 pixel radius for the u,v aperture in deriving visibility amplitudes and phases is optimal for this setup. 

\item Comparing 3\,mm holes vs. 5\,mm holes, the smaller holes lead to higher coherences, and lower noise over the time series, as might be expected due to phase gradients across the holes.

\item Likewise, 1~ms frame time results in much lower rms scatter in the visibility amplitudes and closure phases relative to 3~ms. 

\item We derive a bias in each frame of about 3.7 counts per pixel, and we subtract this bias before analysis. Not subtracting the bias leads to lower coherences by about 2\%. 

\end{itemize}

\subsection{Future Directions}

Following are some recommendations on future experiments that might improve the capabilities of the technique.

\begin{itemize}
    \item Higher physical resolution (in $\mu$m in the source plane): physical resolution at the source (ie. the fringe spacing in units of length) = $\rm L \times (\lambda/B)$, where L is the distance between the mask and the source, $\lambda$ is the observing wavelength, and B is the baseline ($\rm \lambda/B$ is the fringe spacing in radians). Higher physical resolution can be realized in three ways: 
    \begin{itemize}
    \item Shorter wavelength (decrease $\lambda$): change the filter to a shorter wavelength. The question then becomes one of performance of mirrors, lenses, and mirrors, and the source counts. 
    \item Longer baselines on the mask (increase B): this is limited by the light beam foot-print on the mask (Figure~\ref{fig:mask}), which is dictated by the pick-off optics and window in Xanadu.
    \end{itemize}

    \item Measure the aperture illumination, or gains, individually at the start of the experiment, to compare with the fitted gains.  
        
    \item Shorter integration (frame exposure) times could be used to check the coherence. 
    
    \item Vary the electron beam size in real time and verify the measurement method.
    
    \item Precision dark frames: Match the integration times of observing frames before and after. 

    \item  Rotate the mask to fill out the uv coverage. 

\end{itemize}

\subsection{Beyond Gaussianity}

Beyond the Gaussian assumption for synchrotron light sources, there is the broader issue of characterizing the shape of high energy electron beams used across industry and science, and the potential importance of non-Gaussian electron distributions. Examples include high energy electron beam applications in which the beam is shaped to conform to some specific application, such as medical usage to match beam shape to target size, or the recent technique of using transverse resonance island buckets in synchrotron light sources to provide multiple stable orbits winding around the main beam orbit (Goslawski et al. 2017). 

A multi-hole, non-redundant mask and subsequent Fourier imaging analysis, including deconvolution of the point response function (Fourier transform of the u,v sampling), and self-calibration in both phase and amplitude, could be implemented to determine more complex electron beam distributions, without strict {\sl a priori} model assumptions. The process could parallel the ‘hybrid mapping’ process employed in astronomical Very Long Baseline Interferometry, which includes self-calibration and some form of deconvolution (see Pearson \& Readhead 1984).  A requirement of such a process is the need for more measurements (visibilities) than degrees of freedom in the source. For example, a double Gaussian would have 10 degrees of freedom = position (2) and shape (3) of each Gaussian.  We have generated a template for a 7-hole non-redundant mask based on the aperture mask used on the James Webb Space Telescope that fills the light footprint on the mask (Sivarmakarishnan et al. 2024).  The full complex gain self-calibration of amplitudes and phases also represent a precise wavefront sensor, where the phases correspond to the photon path-length through the optical system. 

More generally, characterizing the shape of intense light sources other than synchrotron sources may be possible with this technique, such as in inertial fusion reactors, where the driving lasers need to be highly focused shaped, and the shape of plasma EUV light sources in photo-lithography.

\vskip 0.3in

{\bf Acknowledgments.} The National Radio Astronomy Observatory is a facility of the National Science Foundation operated under cooperative agreement by Associated Universities, Inc.. Patent applied for: UK Patent Application Number 2406928.8, USA Applications No. 63/648,303 (RL 8127.306.USPR) and No. 63/648,284 (RL 8127.035.GBPR).

\clearpage
\newpage

{\bf REFERENCES}

\noindent Born \& Wolf 1999, Principles of Optics, CUP: Cambridge

\noindent Butti, D. et al. 2022, IBIC2022 Krakow, MOP23, p. 88

\noindent Buscher, D. \& Haniff, C. 1993, Journal of the Optical Society of America A, Volume 10, p. 1882

\noindent van Cittert 1934, Physica 1, 201–210 

\noindent Carilli, Nikolic, Thyagarjan 2023, Journal of the Optical Society of America A, Volume 39, 2214

\noindent Cornwell \& Wilkinson 1981,  MNRAS 196, 1067–1086 

\noindent Elleaume, P et al. 1995 J. Synch. Rad., 2, 209

\noindent Golawski. P. et al. 2017, IPAC Copenhagen, p. 3059

\noindent Gonzales-Mejia 2011,  Journal of the Optical Society of America A, Volume 28, p. 6

\noindent Haniff, C. et al. 1989, MNRAS, 241, 51

\noindent Hinkley et al. 2022, Proceedings of the SPIE, Volume 12180, p. 25 

\noindent Jennison 1958 MNRAS 118, 276 

\noindent Kube 2007, Proceedings of DIPAC 2007, Venice, Italy, p. 6

\noindent Labeyrie 1996, A\& AS 118, 517

\noindent Lau et al. 2023, arXiv:2311.15948

\noindent Masaki \& Takano 2003, J. Synch. Radiation, 10, 295

\noindent Michelson 1890, Phil. Mag. 30, 1

\noindent Mitsuhasi 2012, Proceedings of IBIC2012, Tsukuba, Japan, p. 576

\noindent Monnier 2003, Reports on Progress in Physics, Volume 66, Issue 5, p. 789

\noindent Nikolic, B. et al. 2024, Phys. Rev. Accelerator Physics, submitted (arXiv: 2405.12090)

\noindent Novokshonov et al. 2017, Proceedings of IPAC2017, Copenhagen, Denmark, p. 177

\noindent Pearson, T. \& Readhead A. 1984, ARAA, 22, 97

\noindent Readhead \& Wilkinson 1978, ApJ 223, 25

\noindent Schwab 1980, International Optical Computing Conference I, vol. 231 of Society of Photo-Optical Instrumentation Engineers (SPIE) Conference Series W. T. Rhodes, ed., p. 18

\noindent F. Schwab 1981,  NRAO VLA Sci. Memo. 136 

\noindent Sivaramakrishnan et al. 2024, arXiv:2210:17434

\noindent Skopintsev et al. 2014, J. Syncrotron Rad., 21, 722

\noindent Taylor, Carilli, Perley eds. 1999, Synthesis Imaging in Radio Astronomy II, vol. 180 of Astronomical Society of the Pacific Conference Series (San Francisco, Calif. : Astronomical Society of the Pacific)

\noindent Thomson, Moran, Swenson 2023,  Interferometry and Synthesis in Radio Astronomy, 4th Edition (Springer)

\noindent Thyagarajan \& Carilli 2022, PASA, 39, 14

\noindent Torino \& Iriso 2016, Phys Rev Accelerators \& Beams, 19, 122801


\noindent Torino \& Iriso 2015, Proceedings of IBIC2015, Melbourne, Australia, p 428

\noindent Zernike 1938, Physica 5, 785–795 (1938)

\end{document}